\documentclass{article}
\usepackage[letterpaper,top=2cm,bottom=2cm,left=2.5cm,right=2.5cm,marginparwidth=1.75cm]{geometry}
\usepackage[utf8]{inputenc}
\usepackage{amsmath}
\usepackage{amssymb}  
\usepackage{graphicx}
\usepackage[colorlinks=true, allcolors=blue]{hyperref}
\usepackage[affil-it,blocks]{authblk}
\usepackage{siunitx}
\usepackage{ulem}
\usepackage{mathrsfs}
\usepackage{float}
\usepackage{pdfpages}
\usepackage{overpic}
\usepackage{subcaption}               
\usepackage{placeins}                 
\usepackage{orcidlink}                
\usepackage[affil-it,blocks]{authblk} 

\usepackage[sorting=none]{biblatex} 
\title{High precision measurement of phi-nucleon cross section using a tensor polarized deuteron target\\(PAC53)}
\date{\today}

\begin{document}

\author{Mark~M.~Dalton$^{\dagger}$~\orcidlink{0000-0003-2758-6526}} 
\author{Alexandre~Deur$^{*}$~\orcidlink{0000-0001-9204-7559}} 
\author{Chris~D.~Keith$^{*}$~\orcidlink{0000-0001-9204-7559}} 
\affil{Thomas Jefferson National Accelerator Facility, Newport News, VA, USA}

\author{Nadia~Fomin$^{*}$~\orcidlink{0000-0002-4837-3223}} 
\affil{University of Tennessee, Knoxville, TN, USA}

\author{Theory support: Misak~Sargsian~\orcidlink{0000-0003-2762-6305}} 
\affil{Florida International University, Miami, FL, USA}

\date{\small{*: Co-spokesperson, $\dagger$: Contact-spokesperson (dalton@jlab.org)}}

\maketitle


\section*{Executive Summary}

We propose to measure the $\phi$-nucleon cross section $\sigma_{\phi N}$ to solve the longstanding 
puzzle of whether $\sigma_{\phi N} \simeq 10$~mb, as extracted from photoproduction, or $\sim$30~mb, as obtained from nuclear rescattering. CLAS data demonstrated that even precision data for unpolarized $\phi$ photoproduction are insufficient to unambiguously extract $\sigma_{\phi N}$, allowing the possibilities of both $\sigma_{\phi N}$ values. However, as it often happens,  the additional spin degrees of freedom afforded by a polarized target sufficiently constrains the theory to unambiguously provide $\sigma_{\phi N}$.
This will be accomplished with a measurement of the tensor asymmetry $A_{zz}$ in coherent $\phi$ photoproduction from the deuteron, 
$\gamma + d \to \phi + d$.
The same measurement in coherent $\rho$ photoproduction  will allow us to understand the kinematic dependence of the photon longitudinal interaction length for this process, which is necessary to identify unambiguous signal for color transparency. 
This will be the first study of observables in photoproduction from tensor polarized deuterons.

We will use the standard GlueX spectrometer, the Hall D Dynamic Nuclear Polarization (DNP) polarized target, and both a circularly and linearly polarized tagged photon beam. 
The ability to operate the target in a frozen spin mode and produce negative tensor-polarization will decrease the time needed to make the measurements by a factor of more than 2.5.
We request 65 days of beam time which includes commissioning the tensor polarized target.

\tableofcontents

\newpage
\section{Introduction}

In this proposal we describe an experiment to measure the $\phi$-nucleon tensor-polarization cross section to high precision in order to resolve the 50 year old question of its value~\cite{Feynman:1973xc}. 
It will open the door to new studies of the interaction of particles with nuclear matter in a well-controlled environment.
Concurrently, we will also measure various final states in photoproduction from tensor polarized deuterons. 
This may provide a unique new perspective on the deuteron, in particular allowing for the separation of the $S$- and $D$-wave components of the deuteron wave function.

\subsection{Physics of Interest}

Photoproduction of the $\phi$ meson off a tensor-polarized deuteron target can uniquely constrain the $\phi$-nucleon cross section, providing a means to solve the long-standing puzzle of what this cross section is (Section~\ref{sec:physics:phiN}).
The $\phi$-nucleon cross section is a crucial component for modern QCD inspired models.
The $\phi$ meson is essentially a pure $s \bar s$ state~\cite{ParticleDataGroup:2024cfk}. 
The quark-interchange component of the hadron-hadron interaction, including during photoproduction via vector meson dominance (VMD), is thus highly suppressed in the $\phi N$ cross section $\sigma_{\phi N}$. 
Therefore, measuring it allows to isolate the purely gluonic component of the interaction, the Pomeron exchange. 
\begin{figure}[htb]
\center
\includegraphics[width=1.\textwidth]{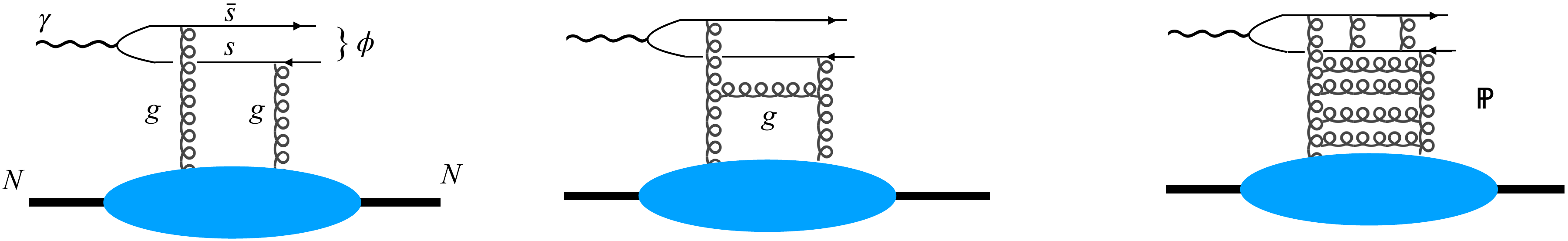}
\caption{\small The $\phi$ meson is largely a purely $s\bar s$ state. Therefore the $\phi$-nucleon interaction is expected to occur at minimum through 2 gluon exchange in the hard regime (left), evolving to pomeron exchange in the softer regime (right).
}
\label{fig:gluon_exchange}
\end{figure}

Yet the determination of this coupling produces diverging values depending on the method used.
A VMD analysis of the forward $\gamma p\to\phi p$ differential cross section produces a value of $\sigma_{\phi N}\eqsim11$ mb~\cite{Sibirtsev:2006yk}.
Meanwhile, the $A$-dependence of the incoherent $\phi$ photo-production and a Glauber-type multiple scattering theory produces higher values,  $\sigma_{\phi N}\sim30$~mb~\cite{Sibirtsev:2006yk,Ishikawa:2004id}.
Accounting for in-medium modification of the $\phi$ does not resolve such strong disagreement~\cite{Cabrera:2003wb,Muhlich:2005kf}.
Most recently, unpolarized scattering $\gamma d\to\phi d$ produced results consistent with both the low and high cross section values, depending on whether VMD is employed or not~\cite{CLAS:2007xhu}.

Measuring the $\phi$-nucleon cross section thus provides a fundamental ingredient needed for nuclear modeling, but also has the ability to shed light on the limits of VMD and the formation of the the $\phi$.
The measurement we propose here provides additional degrees of freedom (compared to unpolarized measurements) that disentangle cross section for the $\phi$, $\sigma_{\phi N}$, from other effects.

Concurrently to $\phi$ production, we will also measure other photoproduced mesons, in particular the $\rho$.
Such a measurement off a tensor-polarized deuteron target can serve as a filter for short-range interactions whose importance in nucleon-nucleon interaction and hadronic structure has long been known~\cite{Frankfurt:1981mk} but has recently come under increased focus thanks to recent JLab results on short-range correlations (SRC)~\cite{Fomin:2017ydn, Arrington:2022sov} and the evolution of our understanding of the EMC effect~\cite{EuropeanMuon:1983wih, Norton:2003cb, Frankfurt:1988nt, Malace:2014uea}.  Short-range interactions can be isolated by studying the $D$-wave of the deuteron because it is highly suppressed at low-momentum due to the orbital angular momentum (OAM) barrier. Indeed, the centrifugal repulsion forces systems with large OAM values ($l=2$ for the $D$-wave) to have large momentum. 
In turn, the $D$-wave is naturally isolated using tensor-polarization, see next section, for which the $\rho$ provides the easiest measurement. 
As will be discussed in more detail in Section~\ref{sec:physics:rho}, short range phenomenology opens a window on the role of relativistic effects in nuclear structures,  is critical to understand SRC observations ({\it e.g.}, the dominance of $np$ SRC pairs over $pp$ and $nn$ pairs~\cite{Tang:2002ww, Subedi:2008zz, Sargsian:2005ru, Piasetzky:2006ai, Schiavilla:2006xx, Arrington:2022sov}) as well as the EMC effect due to evidence that this one is linked to the high-momentum component of the nuclear wavefunction~\cite{Weinstein:2010rt}. Additionally, a detailed understanding of short range phenomenology will most likely be crucial to model extreme nuclear matter states such as neutron star structure~\cite{Frankfurt:2008zv, Li:2019kua, Lu:2021xvj, Hong:2024odm}. Finally, since the $D$-wave characterizes pairs of close nucleons, a reaction with large $D$-wave contribution displays enhanced double scattering effect, in particular in $\rho$ productions~\cite{Frankfurt:1997ss, Frankfurt:1998vx, Freese:2013adl}, which makes it an important tool to study color transparency (CT)~\cite{Bertsch:1981py, Kopeliovich:1981pz, Frankfurt:1992dx, Brodsky:1994kf, Frankfurt:1998vx, Sargsian:2002wc}.

\subsection{Deuteron Structure and Tensor Polarization}

The deuteron is the simplest nuclear system and should be very well known experimentally. However, there remain significant uncertainties for small internucleon distances, or larger nucleon momenta. This dearth of knowledge is particularly unfortunate since these momenta are common for nucleons in heavier nuclei.

\begin{figure}[htb]
  \includegraphics[width=.99\linewidth]{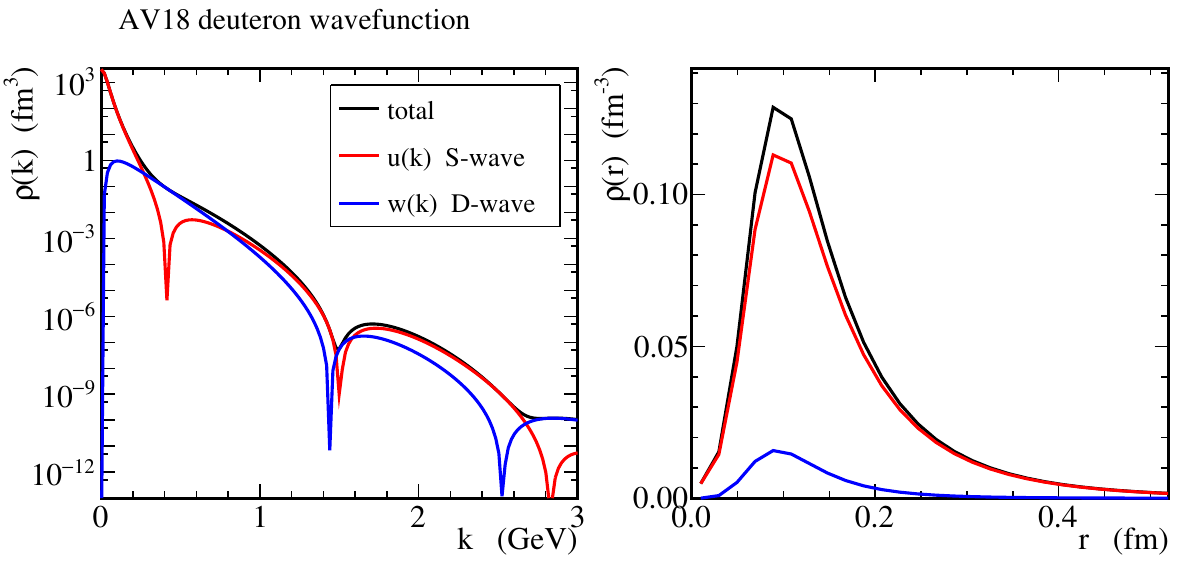}
\caption{Deuteron wavefunction from AV18~\cite{Wiringa:2013ala} showing the $S$- and $D$-wave components and the total.  Left panel in momentum space and right panel in coordinate space. }
\label{fig:D_wavefunction}
\end{figure}

The wavefunction of the deuteron in momentum and coordinate spaces is shown in Fig.~\ref{fig:D_wavefunction} for the AV18 potential~\cite{Wiringa:2013ala}.
One can observe the curious fact that although the $S$-wave is much larger than $D$-wave for all internucleon distances, the $D$-wave dominates in momentum space between 300 and 800 MeV/c. 
In the region just above the 300 MeV/c characteristic Fermi momentum in nuclei, the $S$-wave momentum sign oscillates, making the $D$-wave function in momentum space  several times larger.

The higher momentum components of the wavefunction correspond, via the Fourier transform, to a small coordinate space interval. 
As such, the study of the deuteron in the $D$-wave represents a unique opportunity to study a two-nucleon system in the configuration where momenta are large and the inter-nucleon distances are likely to be small.  

The only known  way  to separate $S$- and $D$-waves is using polarized deuteron beams or targets. 
The wavefunction is relatively well-known and composed of a dominating spherically-symmetric $S$-wave and about 4\% of $D$-wave. The quadrupole shape of the latter implies an angular dependence of the wavefunction that can only be generated by a {\it tensor} component in the nuclear force, {\it i.e.}, a force that depends explicitly on two independent directions. In a deuteron, these directions can only be provided by 
(1) the orientation of the axis between the proton and neutron and 
(2) the orientation of the spin's quantization axis. Therefore, spin degrees of freedom (d.o.f.) are fundamental to the existence of the tensor force and, in turn, of the $D$-wave. 
For example, the transition between the $S$-state of OAM $l=0$ and the $D$-state with $l=2$ occurs when both p and n flip their spins --which become antiparallel to the deuteron spin-- can occur only under the influence of the tensor force since neither scalar nor vector can induce $\Delta l=2$ transitions. 
Appendix~\ref{sec:add_physics} briefly describes some ways that this experiment might be sensitive to deuteron structure, although not the focus of this proposal.

\begin{figure}[htb]
\center
\includegraphics[width=0.6\textwidth]{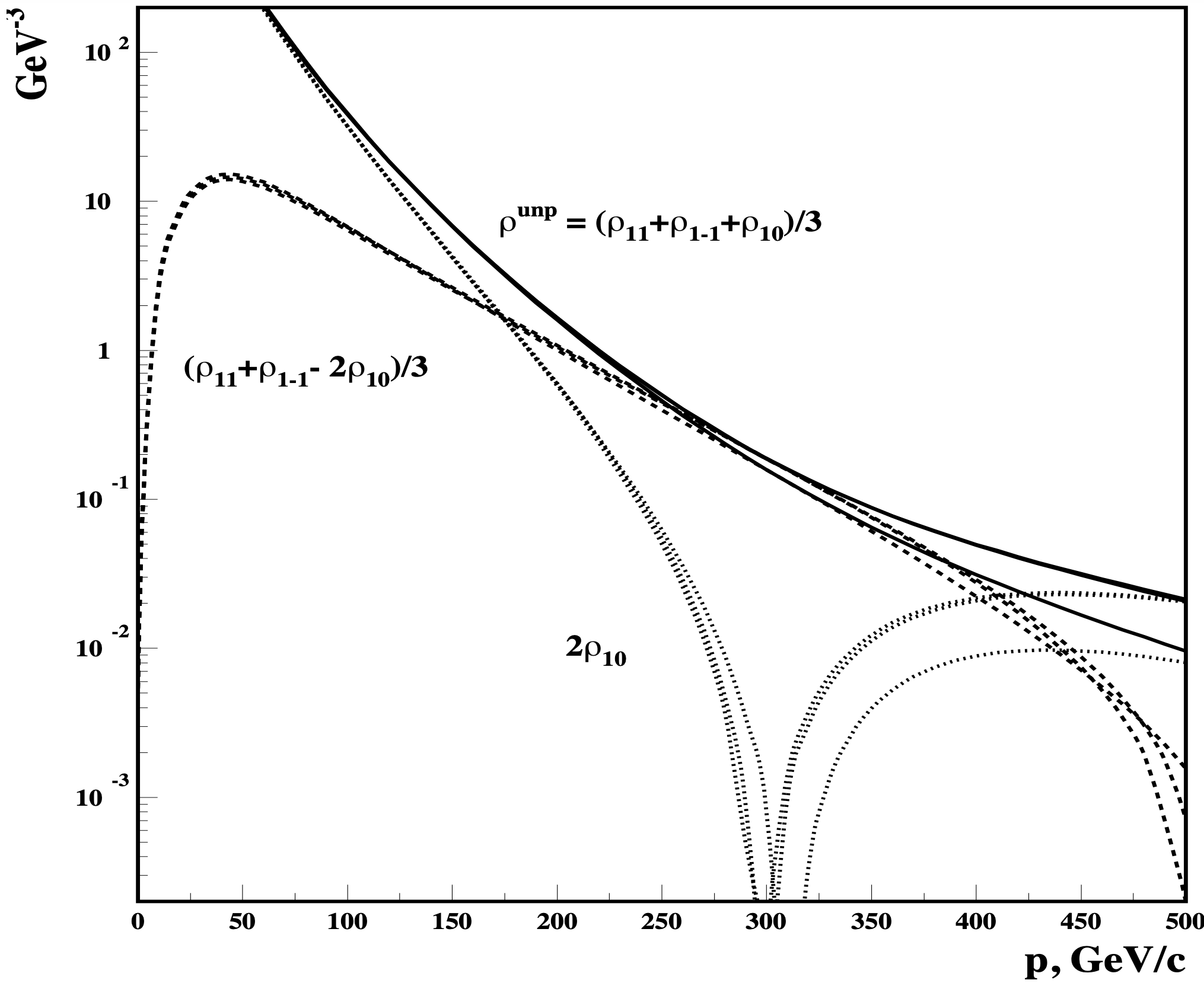}
\caption{\small The deuteron density matrix $\rho_{1,m}$ as a function of the relative momentum $p$ between proton and neutron, for 3 different model deuteron wavefunctions. The subscripts $1,m$ indicate the deuteron spin and its projection on the quantization axis, respectively. The solid curve shows the unpolarized case, $\rho^{\rm unp}$. The dashed curve is for ``tensor-polarized" case and the dotted curve the $m=0$ projection $\rho_{10}$.  The latter shows the node of the deuteron's charge form factor.  Note that wavefunction models do not agree on the location of the minimum.
Figure from Ref.~\cite{Sargsian:2014fla}}
\label{fig:Sargsian2014_fig1}
\end{figure}
Figure~\ref{fig:Sargsian2014_fig1} shows the deuteron density matrix $\rho_{1,m}$ as a function of the relative momentum $p$ between proton and neutron.
As explained in Ref.~\cite{Sargsian:2014fla}, constructing observables of the form 
\begin{equation}\label{eq:tensor_density}
   \rho^{20} = \frac{1}{3}(\rho^{11}+\rho^{1-1}-2\rho^{10}),
\end{equation}
where $\rho^{20}$ is the $D$-wave component and transforms like a tensor, provides a unique possibility for studying the NN strong interaction at short space-time separations.  
(In Ref.~\cite{Sargsian:2014fla} this combination is referred to as ``Tensor Polarized", here we denote it $\rho^{20}$ to avoid ambiguity with target nomenclature.) Experimentally, the combination~(\ref{eq:tensor_density}) can be achieved by determining individually the spin-separated observables  $\rho^{11}$, $\rho^{1-1}$, and $\rho^{10}$ and then combining them according to Eq.~(\ref{eq:tensor_density}).
Denoting $u(k)$ and $w(k)$ the $S$ and $D$ partial waves of the deuteron wavefunction respectively, the $\rho^{20}$ density matrix depends only on the terms proportional to $u(p)w(p)$ and $w(p)^2$~\cite{Sargsian:2014fla}, which greatly diminishes the low momentum strength, Fig.~\ref{fig:Sargsian2014_fig1}.

The new sensitivity to meson rescattering with a tensor-polarized target may be thought of in a geometric sense.  
The $m=\pm1$ and $m=0$ spin states are understood to have different spatial density distributions and can be visualized as having a ``peanut" and ``donut" shape respectively.
The donut shape of the $m=0$ is related to the node in the charge form factor (observable in the $m=0$ density matrix ($\rho_{10}$) of Fig.~\ref{fig:Sargsian2014_fig1}.
The location and depth of the resulting node in the $m=0$ cross section depends very sensitively on double scattering within the deuteron.

\begin{figure}[htb]
\center
\includegraphics[width=0.6\textwidth]{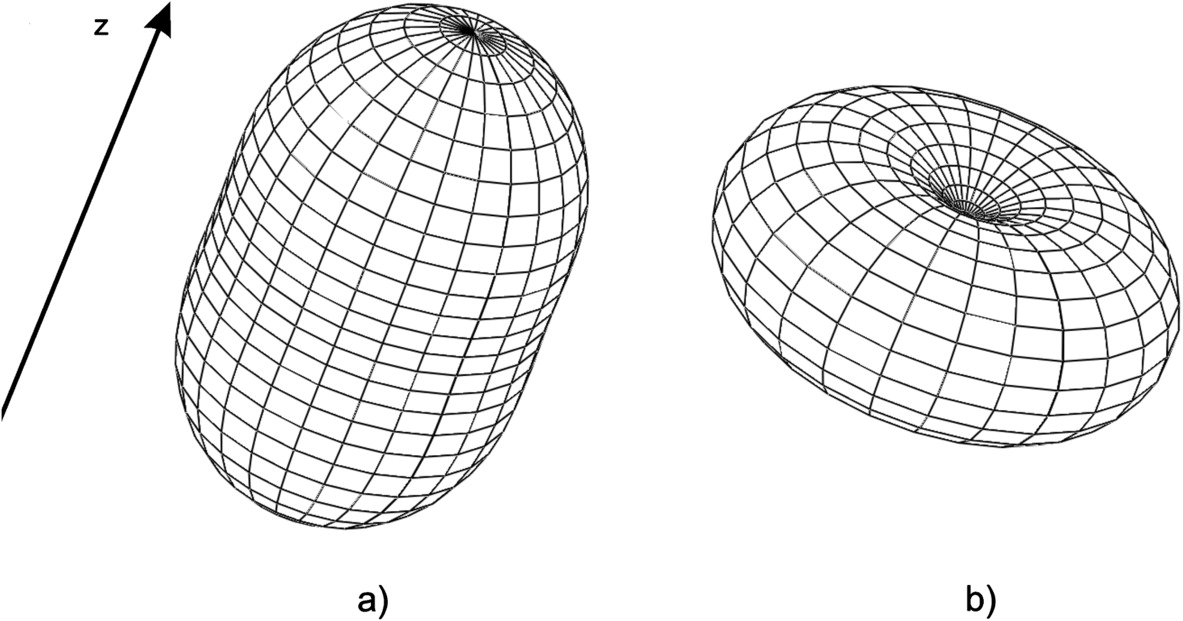}
\caption{\small Visualizations of the nucleon densities of the deuteron in its two spin projections a) $m=\pm1$ and b) $m=0$, with their respective ``peanut" and ``donut" shapes.
Figure from Ref.~\cite{Baryshevsky:2010zz}}
\label{fig:deuteron_shapes}
\end{figure}

Tensor polarization, denoted here as $Q$, is one of two orientation parameters that are necessary to describe an ensemble of identical spin-1 particles in a magnetic field with cylindrical symmetry.  The other parameter is the familiar vector polarization, $P$.  These are defined as: 
\begin{eqnarray}
    P &=& N_{+} - N_{-}~,  \label{eq:Pz} \\
     Q&=& (N_+ -N_0) - (N_0 - N_-) = 1 - 3N_0 ~. \label{eq:Pzz}
\end{eqnarray}
Here $N_{+}$, $N_{-}$ and $N_{0}$ are the number of deuterons with their spins in the $m = +1,-1,0$ projections respectively, normalized such that
\begin{equation}
\label{eq:norm}
    N_{+} + N_{-} + N_{0} = 1~.
\end{equation}
Written in terms of the polarizations, the fractional populations are:
\begin{eqnarray}
\label{eq:N+}
    N_{+} &=& \frac{1}{3} + \frac{P}{2} +\frac{Q}{6}~, \\
\label{eq:N0}
    N_{0} &=& \frac{1}{3}(1 - Q)~, \\
\label{eq:N-}
    N_{-} &=& \frac{1}{3} - \frac{P}{2} +\frac{Q}{6}~.
\end{eqnarray}
The tensor polarization is thus a measure of the $N_{0}$ population.  $Q>0$ indicates a relative depletion of the $m = 0$ projection, while $Q<0$ indicates its enrichment.  
The ability to tensor-polarize a deuteron target is equivalent to being able to spin-separate the initial deuteron state into the spin states $m=0,\pm1$.

\begin{figure}[htb]
\center
\includegraphics[width=0.4\textwidth]{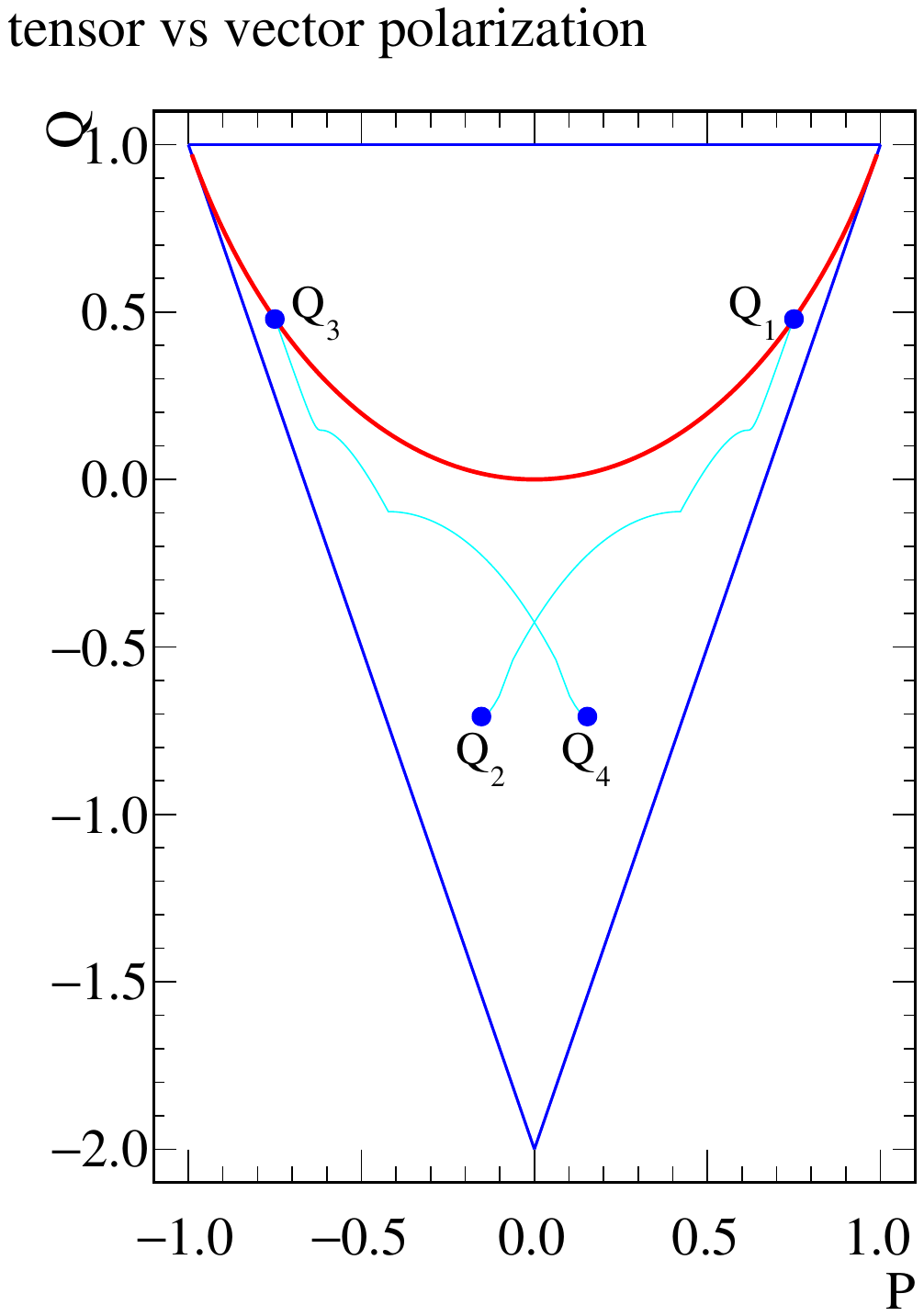}
\caption{\small Plot of the tensor versus vector polarizations.  Physical states lie within the triangle.  The red curve shows those states that can be achieved by the DNP technique and obeying Eq.~(\ref{eq:PzzfromPz}).  $Q_{1,2,3,4}$ are the spin states that we intend to employ.
}
\label{fig:triangle}
\end{figure}

In systems where the three projections are populated according to a Boltzmann distribution, then 
\begin{equation}
\label{eq:PzzfromPz}
    Q = 2 - \sqrt{4 - 3P^2}.
\end{equation}
This relation holds true when an ensemble of spin-1 nuclei is in thermal equilibrium at a uniform temperature $T$.  It remains true in many solid systems in which the vector polarization has been enhanced using Dynamic Nuclear Polarization (DNP).  In fact, DNP is frequently described as the cooling of a spin system to a ``spin temperature'' $T_s$ (positive or negative) far below the lattice temperature.  In these circumstances, Eq.~(\ref{eq:PzzfromPz}) implies that a negative tensor polarization cannot be obtained from dynamic polarization alone.  
The technique that we will use to significantly alter $Q$ from this ``Boltzmann restriction'' is Adiabatic Fast Passage (AFP).  Unlike other techniques, AFP can produce high degrees of positive or negative $Q$ and does not require a concentration of protons within the sample.  
Figure~\ref{fig:triangle} shows the tensor versus vector polarization and the spin states $Q_{1,2,3,4}$ of the target. We intend to take data for an equal amount of time in each of 4 states, see Sec.~\ref{sec:target} for details.
Cycling through the states will be done using AFP manipulation,
$Q_{1} \xrightarrow{\text{AFP}}Q_{2}\xrightarrow{\text{AFP}}Q_{3}\xrightarrow{\text{AFP}}Q_{4}\xrightarrow{\text{AFP}}Q_{1}$.
The line connecting $Q_{1} $ with $Q_{2}$ and $Q_{3} $ with $Q_{4}$ is a partial AFP sweep over the NMR line.

For 4 measurements $\sigma_{1,2,3,4}$ taken with polarizations $Q_{1,2,3,4}$ shown in Fig.~\ref{fig:triangle}, the tensor asymmetry is given by 
\begin{equation}
    \label{eq:Azz_2qvals}
A_{zz} = 2\frac{\sigma_+-\sigma_-}{Q_{+}\sigma_--Q_{-}\sigma_+},
\end{equation}
where $\sigma_{+} = (\sigma_{1} + \sigma_{3})/2$ and $\sigma_{-} = (\sigma_{2} + \sigma_{4})/2$ are the average cross sections for the positive and negative tensor polarization settings respectively, constructed to cancel any vector polarization component. Here, $Q_{+} = Q_{1} = Q_{3}$ and $Q_{-} = Q_{2} = Q_{4}$ are the values of the positive and negative tensor polarizations respectively.

\subsection{Summary of existing efforts using tensor-polarized deuterons}

Tensor-polarized deuterons have been employed in several experimental programs to probe the spin structure and electromagnetic properties of the deuteron, as well as to test nuclear interaction models. The HERMES experiment at DESY \cite{HERMES:2005pon} performed deep inelastic scattering measurements using a tensor-polarized internal gas target to extract the deuteron's tensor structure function $b_1$. At MIT-Bates, the BLAST experiment, also using a tensor-polarized internal gas target, studied the structure of the deuteron, including form factors and tensor analyzing powers \cite{DeGrush:2017azm, Zhang:2011zu}. At JINR (Dubna), tensor-polarized deuteron beams have been used to measure the analyzing powers and spin observables in $dp$ scattering and breakup reactions \cite{Ladygin:2011zz}. COSY (FZ Jülich) and RIKEN have also employed tensor-polarized deuteron beams to investigate three-nucleon forces and short-range correlations via spin-dependent measurements in $dp$ reactions \cite{2006APS..APRC14006M, Sekiguchi:2004yb, Ladygin:2004pd}. Additionally, the KVI facility in the Netherlands has contributed significantly through measurements of polarization observables in $dp$ breakup reactions, providing constraints on three-nucleon force models \cite{Stephan:2010zz, Stephan:2007zza}.   
At Jefferson Lab in Hall C, $t_{20}$ was measured in elastic $ed$ scattering using recoil deuteron polarimetry~\cite{JLABt20:2000uor}. 
Further measurements of $t_{20}$~\cite{PR12-15-005_Long} and $b_1$~\cite{PR12-13-011_Slifer} are planned using a tensor-polarized solid target at JLab~\cite{Poudel:2025nof}.

The experiment proposed here differs from these efforts in that it will involve high polarized-luminosity together with large solid angle detection with exclusive capability. This will allow for unprecedented precision and accuracy and, above all, make a large number of observables accessible.

\FloatBarrier
\section{Physics Motivation \label{sec:physics}}

In this section, we describe how measurement of the tensor asymmetry $A_{zz}$ in coherent vector meson photoproduction allows to study the interaction of particles with nuclear matter in a well-controlled environment.  We focus on the $\phi$ as the most puzzling and the $\rho$ as the most precise.

\subsection{The $\phi$-meson--nucleon scattering cross section \label{sec:physics:phiN}}

According to the OZI rule~\cite{Okubo:1963fa, Zweig:1964jf, Iizuka:1966fk}, the total $\phi N$ cross section, $\sigma_{\phi N}$, should be small since the $\phi$ meson consists of almost pure $s\bar{s}$.
The attractive QCD van der Waals force may be an important part of the $\phi N$ interaction and a bound $\phi N$ state may be possible in some reactions.~\cite{Gao:2000az}
This is apparently observed: measured values of $\sigma_{\phi N}$ are around 10~mb, notably smaller than other meson-nucleon total cross sections $\sigma_{\omega N}$, $\sigma_{\rho N}$, and $\sigma_{\eta N}$, which are are $\sim 30$~mb~\cite{Bauer:1977iq,Effenberger:1999ay}.
However, to confuse matters, measurements of $\sigma_{\phi N}$ in nuclear matter yield about 30~mb, consistent with the other meson-nucleon total cross sections.

The $\phi N$ cross section may be determined using a number of methods, which should all agree.
Exclusive $\phi$ photoproduction analyzed with VMD~\cite{Sibirtsev:2006yk} yields $\simeq 10$ mb, as does $\pi N$ and $KN$ data interpreted in the constituent quark model~\cite{Lipkin:1966zzc}.
For coherent nuclear photoproduction of the $\phi$, $\sigma_{\phi N}$ is determined using the $t$-dependence of  the elementary $\gamma N \to \phi N$ photoproduction amplitude, the nuclear form factor and constraints on $\alpha_\phi$, the ratio of the real to imaginary part of the forward scattering amplitude.
The $\sim10$~mb value is well reproduced within a single channel optical model~\cite{Sibirtsev:2006yk}. 

For incoherent nuclear photoproduction of the $\phi$, $\sigma_{\phi N}$ is determined from the $A$-dependence of the $\phi$ yield and a Glauber-type multiple scattering theory.
With a very small $\sigma_{\phi N}$, the incoherent $\phi$ photoproduction cross section from a nucleus would be approximately proportional to the target mass number, $A$.
As $\sigma_{\phi N}$ increases, the exponent $\alpha$ on $\sigma_A\propto A^\alpha$ decreases below unity.
An anomalous $A$-dependence is observed for the $\phi$~\cite{Cabrera:2003wb,Ishikawa:2004id,Muhlich:2005kf}, while the $A$-dependence of the $\omega$-meson production both in $\gamma A$ and $pA$ interactions is well understood~\cite{Sibirtsev:2006yk}.
The peculiar $A$-dependence implies that the $\phi N$ interaction is stronger than theoretical expectations.  On possibility is that the $\phi$ properties are modified in the nuclear medium~\cite{Cabrera:2003wb,Sibirtsev:2006yk}.
However, this not supported by the fact that the mass and width of the $\phi$ meson observed in the $K^{+}K^{-}$ invariant mass spectrum are consistent with those of the free $\phi$ meson~\cite{Ishikawa:2004id}.  
Also, it would be peculiar that only the $\phi$ structure is modified so extensively that it increases its cross-section by a factor of 3, while such nuclear modifications is not observed for other mesons. 
A different explanation that has been proposed for this anomaly is the the excitation of a cryptoexotic $B_\phi$-baryon~\cite{Sibirtsev:2006yk}.

Yet another potential explanation for these observations is that the $\phi$ is initially produced in a color singlet, small transverse size configuration or point-like configuration (PLC) which then expands over its formation length to become the fully interacting $\phi$, an eigenstate of the QCD Hamiltonian.
In this way exclusive photoproduction from the proton naturally has a small cross section while rescattering in a nuclear environment occurs with a much larger cross section.
This is a similar concept to that of color transparency (CT)~\cite{Bertsch:1981py, Kopeliovich:1981pz, Frankfurt:1992dx, Brodsky:1994kf, Frankfurt:1998vx, Sargsian:2002wc}, except instead of propagating through the nucleus with less interaction than expected, here the interaction is larger because it is compared to the PLC small cross-section.
Color transparency is well established at high energies and low Bjorken-$x$, where a high momentum-transfer preferentially selects the PLC which then moves with high momentum through the nucleus.
An active research program is attempting to learn the limits in energy and momentum for observation of CT, and JLab experiments have shown CT at intermediate energies for $\pi$~\cite{Clasie:2007aa} and $\rho$~\cite{CLAS:2012tlh} electroproduction. 
CT effects are $<20\%$ effect at JLab energies compared to the factor 3 in the $\sigma_{\phi N}$, providing a strong motivation to explore this interesting possibility.
The interpretation of the 10~mb and 30~mb values of $\sigma_{\phi N}$ in terms of the evolution of the PLC explains all current observations and is therefore very compelling. However, this approach does not account for the OZI rule, which is well verified and solidly grounded in QCD. On one hand, the OZI rule appears obligatory, explains the 10~mb value of $\sigma_{\phi N}$ but leaves the 30~mb value to be explained through unconventional and very large medium modifications, or through exotic mechanisms. On the other hand, the PLC evolution provides a compelling explanation for both $\sigma_{\phi N}$ values but ignores the robust OZI rule expectation. This state of affairs clearly calls for new types of constraints from data, such as those offered by tensor observables.

The $\phi N$ cross section $\sigma_{\phi N}$ can be measured using double-scattering from the deuteron in the reaction $\gamma +\vec{d}\to \phi + d$. The deuteron is an ideal system to study the possibility of an initial PLC for the $\phi$  because after production on the first nucleon, the rescattering is limited to a single nucleon, see Fig.~\ref{fig:Frank1}{\color{blue}b}. The initial $\phi$ production from the first scattering is well understood from the proton data and the deuteron wavefunction, well known for low momenta, mitigates uncertainties that are present in heavier nuclei. However, unpolarized data from the deuteron turned out to have a particular correlation between the the $\phi N$ cross section and the $t-$slope of the rescattering which prevented unambiguous extraction of either~\cite{CLAS:2007xhu}. As will be shown, tensor polarization of the Deuteron lifts this degeneracy and provides a clear extraction of both quantities.

\subsubsection{Model of coherent photoproduction of vector mesons from deuterium \label{sec:physics:model}}

\begin{figure}[htb]
\includegraphics[width=0.99\textwidth]{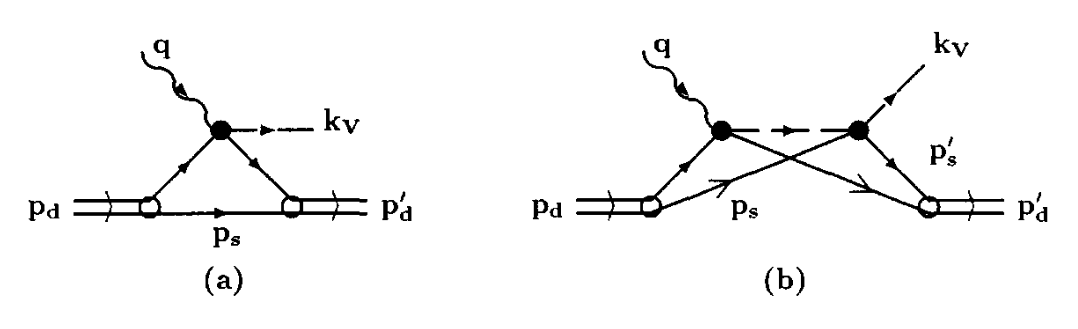}
\caption{\small Vector meson photoproduction off the deuteron in the coherent case, {\it i.e.}, with the deuteron remaining intact after its interaction with the photon. The left diagram depicts single scattering and the right, double  scattering. Figure from Ref.~\cite{Frankfurt:1997ss}.}
\label{fig:Frank1}
\end{figure}

\begin{figure}[htb]
\begin{subfigure}{.33\textwidth}
  \centering
  \includegraphics[width=.99\linewidth]{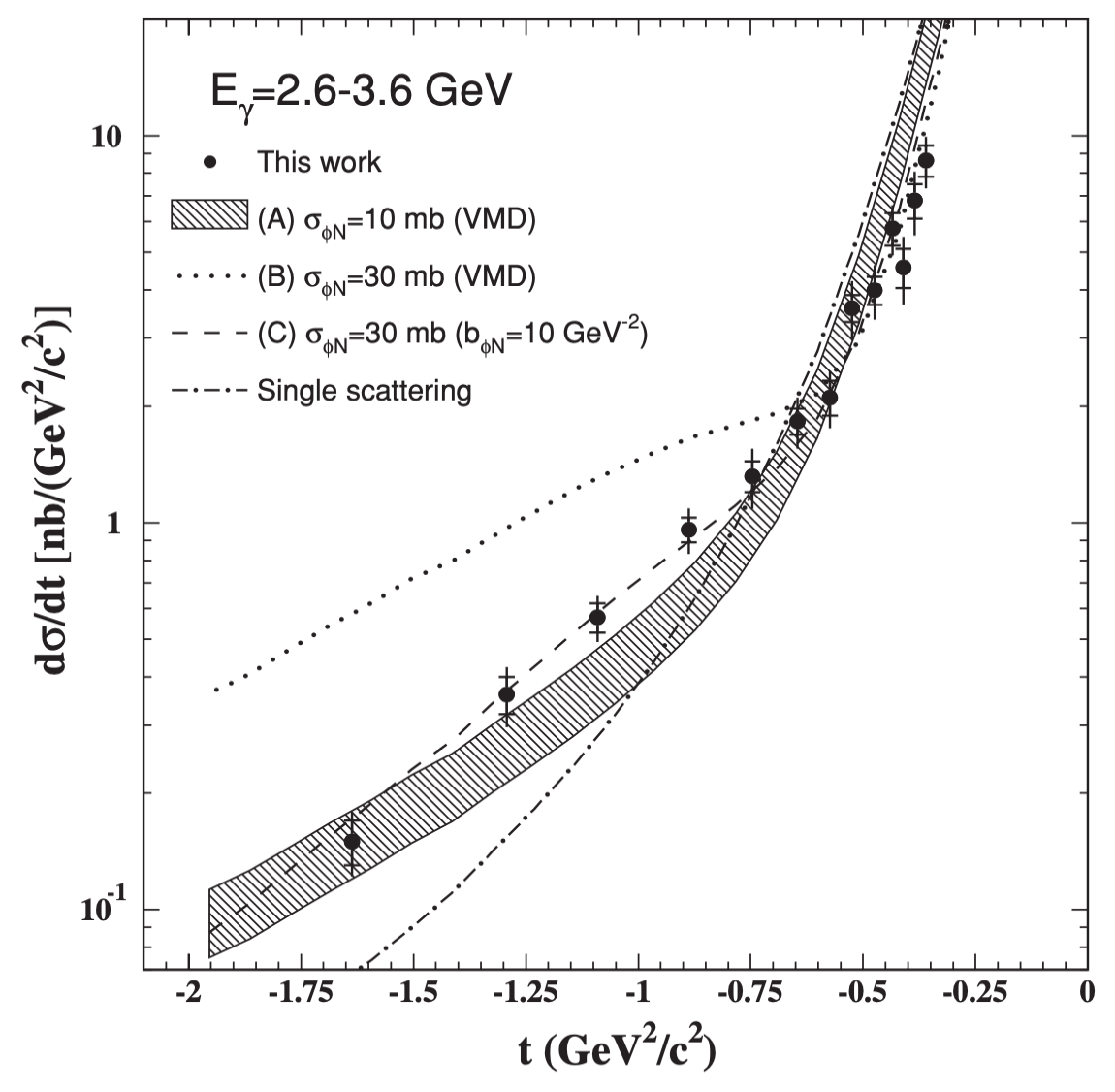}
\end{subfigure}%
\begin{subfigure}{.33\textwidth}
  \centering
  \includegraphics[width=.99\linewidth]{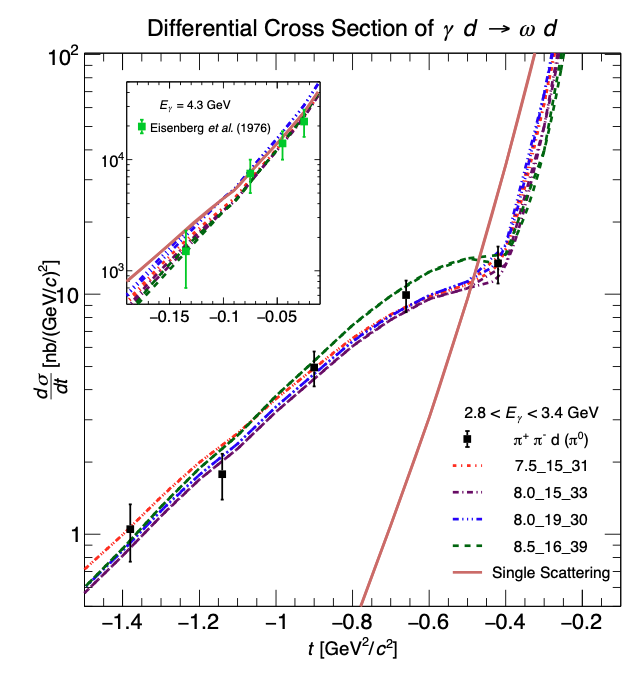}
\end{subfigure}
\begin{subfigure}{.33\textwidth}
  \centering
  \includegraphics[width=.99\linewidth]{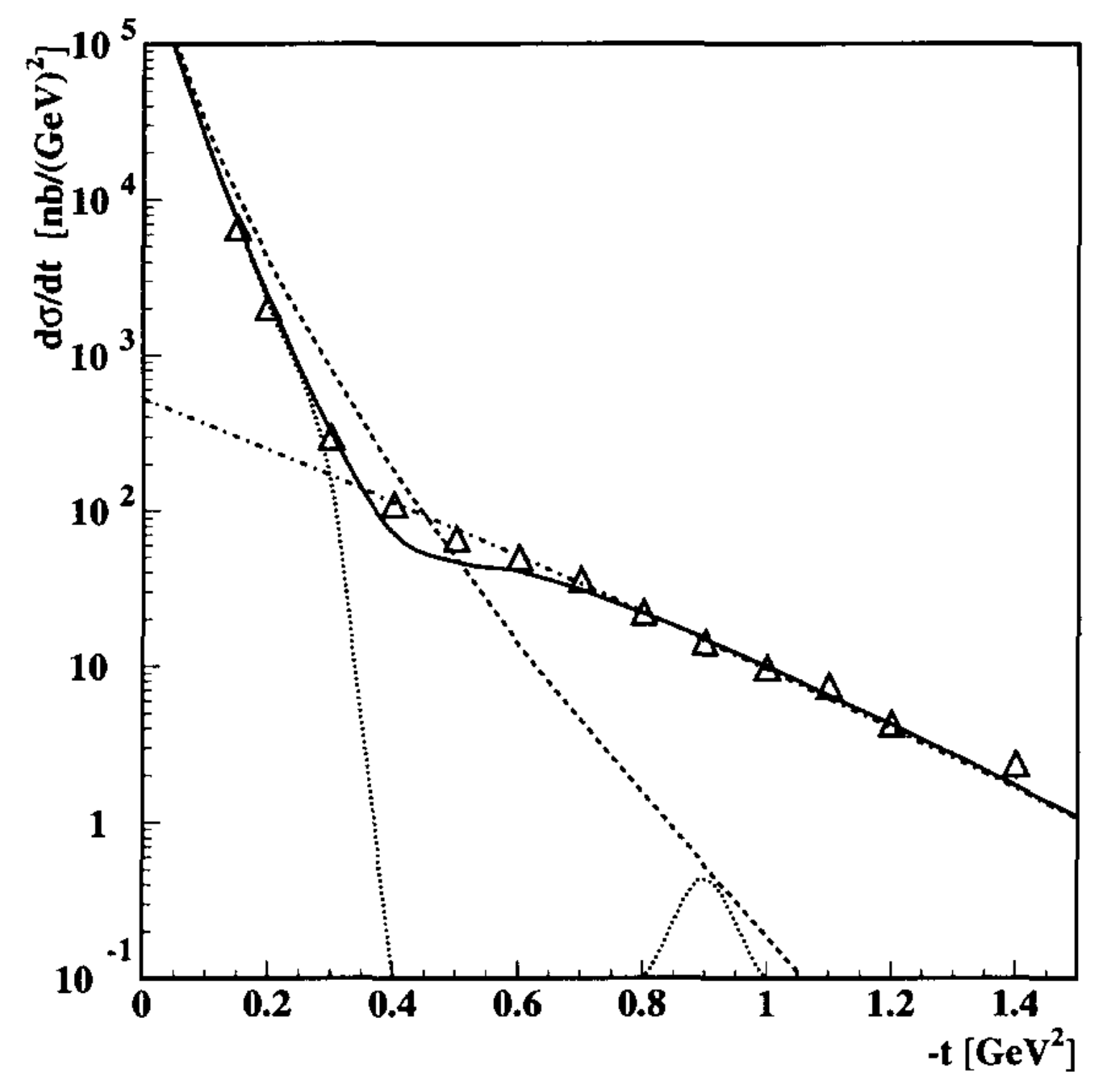}
\end{subfigure}
\caption{
Left: Coherent  $\phi$ photoproduction from deuteron, Ref.~\cite{CLAS:2007xhu}.
Middle: Coherent $\omega$ photoproduction from deuteron, Ref.~\cite{CLAS:2018avi}.
Right: Coherent $\rho$ photoproduction from deuteron at 12 GeV, Ref.~\cite{Anderson:1971ar}.
The data are well described by the calculation of Ref.~\cite{Frankfurt:1997ss}.
}
\label{fig:coherent_unpol}
\end{figure}

Frankfurt {\it et al.}~\cite{Frankfurt:1997ss} developed a theoretical framework to describe coherent photo- and electroproduction from deuterons at intermediate and high energies. 
It is based on vector meson dominance and the eikonal formalism extended to high energy by accounting for recoil effect and the relativistic nature of the deuteron structure. 
The amplitude for the process $\gamma^{(*)} D \to V D$, where $V$ is the vector meson, is derived in the impulse approximation (IA), with corrections from final-state interactions (FSI) between the outgoing meson and the nucleons. 
The full amplitude includes single scattering (Fig.~\ref{fig:Frank1}{\color{blue}a}), double scattering (Fig.~\ref{fig:Frank1}{\color{blue}b}) and their interference. 
The amplitude is integrated with the deuteron wave function (Ref.~\cite{Frankfurt:1997ss} used the Paris potential) to factor in the choice of a coherent reaction, viz with the deuteron remaining intact. In doing so, the eikonal approximation is used when accounting for the FSI in order to extend the standard Glauber approximation by incorporating recoil and relativistic effects. 
This allows accurate modeling of rescattering even at moderately high energies, where Glauber theory would otherwise break down. \\
We consider the description~\cite{Frankfurt:1997ss} reliable since it consistently includes FSI (eikonal approximation), uses a realistic deuteron wave function, and above all, reproduces available experimental data on vector meson production from SLAC~\cite{Anderson:1971ar}, and JLab~\cite{CLAS:2007xhu,CLAS:2018avi}, see Fig.~\ref{fig:coherent_unpol}.

This approach to extract $\sigma_{\phi N}$ was first used in CLAS with unpolarized deuteron and photon energies ranging from 1.6 to 3.6 GeV ($-2.00 \leq t \leq -0.35$ GeV$^2$)~\cite{CLAS:2007xhu}. 
However, unpolarized data were insufficient to solve the $\sigma_{\phi N}$ puzzle: the CLAS data yield either $\sigma_{\phi N} \simeq 10$ mb when analyzed within strictly VMD, (with the slope of the diffractive amplitude $b_{\phi N}=6$ (GeV/c)$^{-2}$), or $\sigma_{\phi N} \simeq 30$ mb, (with $b_{\phi N}=10$ (GeV/c)$^{-2}$), Fig.~\ref{fig:coherent_unpol} Left.
A larger $\sigma_{\phi N}$ than the VMD prediction is possible if a larger $t$-slope parameter for the $\phi N$ interaction is assumed.

The same technique was again used in CLAS to obtain $\sigma_{\omega N}$, using  unpolarized deuteron and photon energies ranging from 2.8 to 3.4 GeV ($-1.4 \leq t \leq -0.4$ GeV$^2$)~\cite{CLAS:2018avi}.  Again, there is a strong coupling between $b_{\omega N}$ and $\sigma_{\omega N}$.
A scan of the parameters found that a description with $b_{\omega N}=7.5$ (GeV/c)$^{-2}$ and $\sigma_{\omega N}=31$ mb had similar ability to describe the data as $b_{\omega N}=9.0$ (GeV/c)$^{-2}$ and $\sigma_{\omega N}=39$~mb.

\subsubsection{Use of tensor polarization \label{sec:physics:tensor}}

The method outlined here using tensor polarization will unambiguously measure $\sigma_{\phi N}$ at lower $-t$, with higher statistical precision, and significantly better sensitivity to $\sigma_{\phi N}$.
When combined with unpolarized data, it therefore also offer a very sensitive test of VMD for the $\phi$ at these energies.

We use an updated version of the calculation from the authors~\cite{Misak-pers-com} to compute the differential cross section over the energy range in Hall D for the 3 spin states, and to determine the  tensor asymmetry from $A_{zz}=\frac{\sigma_{+1}+\sigma_{-1}-2\sigma_{0}}{\sigma_{+1}+\sigma_{-1}+\sigma_{0}}$.
 
\begin{figure}[htb]
\center
\includegraphics[width=0.99\textwidth]{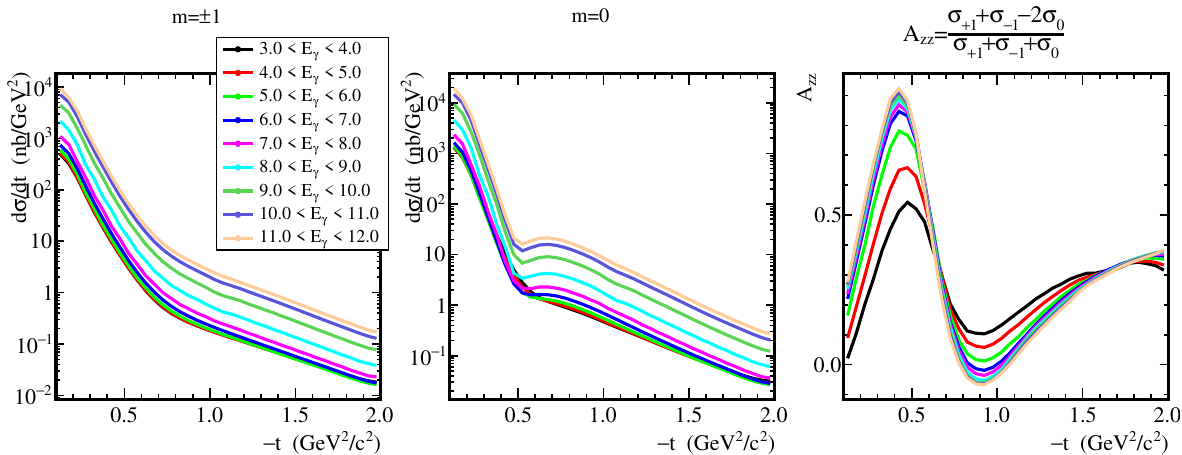}
\caption{\small The differential cross section  (left panel: $m=\pm1$ spin projections, central panel: $m=0$.) 
 and tensor asymmetry $A_{zz}$ (right panel) versus $-t$ for $\gamma d \to \phi d$ for several photon beam energies. 
The calculation uses the updated model of Ref.~\cite{Frankfurt:1997ss} and a $\phi N$ cross section of $\sigma_{\phi N}=10$\,mb with slope of the diffractive amplitude $b_{\phi N}=6$ (GeV/c)$^{-2}$.}
\label{fig:Azz_phi10}
\end{figure}

\begin{figure}[htb]
\center
\includegraphics[width=0.99\textwidth]{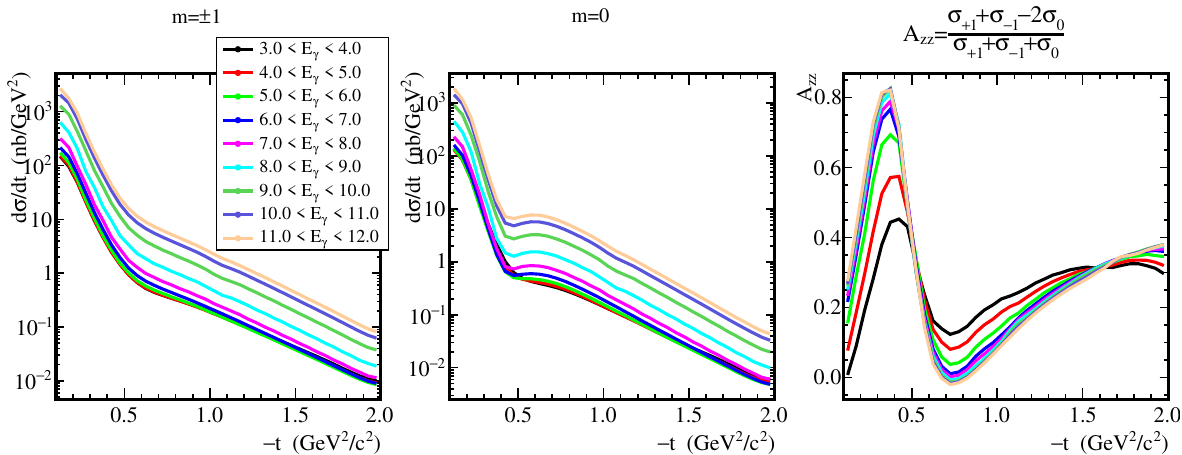}
\caption{\small Same as Fig.~\ref{fig:Azz_phi10} but for $\sigma_{\phi N}=30$\,mb and $b_{\phi N}=10$ (GeV/c)$^{-2}$.}
\label{fig:Azz_phi30b10}
\end{figure}

This is done for the two pairs of parameters considered in Ref.~\cite{CLAS:2007xhu}, namely either for a $\phi N$ cross section of $\sigma_{\phi N}=10$\,mb and slope of $b_{\phi N}=6$ (GeV/c)$^{-2}$ in Fig.~\ref{fig:Azz_phi10}, 
or for $\sigma_{\phi N}=30$\,mb $b_{\phi N}=10$ (GeV/c)$^{-2}$ in Fig.~\ref{fig:Azz_phi30b10}.  
The $m=0$ distribution has a node in the differential cross section that appears at $-t\approx0.4$~GeV$^2$/c$^2$, while the $m=\pm1$ distribution does not.
This leads to a large tensor asymmetry $A_{zz}$, which peaks at the location of this node with an energy dependent amplitude.
Comparing Figs.~\ref{fig:Azz_phi10} and \ref{fig:Azz_phi30b10}, the larger $\sigma_{\phi N}$ amplifies the double-scattering contribution, which increases the ($-t>0.5$ GeV$^{2}/$c$^{2}$) component for all spin states. 
This shifts the node for $m=0$ to lower $-t$, leading to a markedly different shape for $A_{zz}$ that can be exploited to determine the value of $\sigma_{\phi N}$.

We expect and effects of rescattering or final state interactions to be negligible for coherent vector meson production. 
The model of Ref.~\cite{Frankfurt:1997ss} provides an excellent description of unpolarized data for coherent $\rho$~\cite{Anderson:1971ar}, $\omega$~\cite{CLAS:2018avi}, and $\phi$~\cite{CLAS:2007xhu}  photoproduction from deuterons in our energy range.  This can be seen in Fig.~\ref{fig:coherent_unpol}. 
Reference~\cite{Kim:2021adl} studies $\phi$ photoproduction using a Hamiltonian including  Pomeron exchange, meson exchange, $\phi$ radiation, and nucleon resonance excitation mechanisms.  Possible further $\phi N$ interactions are modeled in a comprehensive way using direct $\phi N$ coupling, gluon-exchange and box diagrams from $\pi N$, $\rho N$, $K \Lambda$ and $K \Sigma$ exchange.  This model is then fit to existing  $\gamma p\to \phi p$ data and extended to  $\gamma ^4\textrm{He}\to \phi ^4\textrm{He}$ using the DWIA.  The conclusion is that $\phi N$ rescattering is ``weak" for $p$ and ``negligible" for $^4$He due to being further suppressed by the nuclear form factor.  As the deuteron is weakly bound we can expect it to be similarly small.
Rogers {\it et al.}~\cite{Rogers:2005bt} performs a similar calculation to Ref.~\cite{Frankfurt:1997ss} and uses direct numerical method to account for the Fermi motion of the deuteron.  
The effect of Fermi motion tends to increase the cross section, becoming non-negligible near threshold at high $-t$, but   does not qualitatively impact the conclusions presented here.

In addition to measuring the $\phi$ we will measure $\gamma +\vec{d}\to \rho + d$ and $\gamma +\vec{d}\to \omega + d$.
These data will have higher statistical precision and  parameters for their interaction are better known.
This will allow us to verify that the models we will use for the $\phi$ incorporate all the necessary physics in other cases.

\subsection{Coherent $\rho$-photoproduction \label{sec:physics:rho}}

As previously explained, tensor polarization can serve as a filter to isolate the deuteron $D$-wave and thereby study the short-range interaction contributing to the deuteron structure, and $\rho$-photoproduction offers the easiest reaction for such study due to its large cross section and ease of identification. 

Since the $D$-wave characterizes pairs of close nucleons, a reaction with large $D$-wave contribution displays enhanced FSI, in particular in meson production~\cite{Frankfurt:1997ss, Frankfurt:1998vx, Freese:2013adl}. 
Studying the evolution of FSI in reactions particularly sensitive to them is the best setting to identify Color Transparency (CT)~\cite{Bertsch:1981py, Kopeliovich:1981pz, Frankfurt:1992dx, Brodsky:1994kf, Frankfurt:1998vx, Sargsian:2002wc} since its onset is characterized by the vanishing of the FSI~\cite{Frankfurt:1998vx, Sargsian:2014fla}. 
At JLab energies can be done by studying the survival probability in electroproduction as a function of $Q^2$.
CT touches a central question in hadronic physics: how does hadronization occur, i.e. how a temporarily isolated quark develops into a full-fledged hadron via a small-sized color singlet state (pre-hadron).  
The process is thought to be the origin of CT, the small size of the pre-hadron reducing its FSI with surrounding nuclear matter.
Nuclei can serve as a laboratory to study the space-time evolution of the pre-hadron since its subsequent interactions with surrounding nucleons will evolve as its structure itself evolves into that of a larger hadron~\cite{Raufeisen:2003sd}.  This idea has been fruitful and employed to study the coherence length of virtual photons~\cite{HERMES:1998ajz}, the evolution of the pre-hadrons involved in CT~\cite{HERMES:2002tmh}, and shadowing in deep inelastic lepton-nucleus scattering in the low $x$-Bjorken regime~\cite{NewMuon:1996yuf}.
As the simplest bound state of nucleons, the deuteron is especially important for such studies~\cite{Boeglin:2015cha}. Its wavefunction is well-known and the fact that there are only two nucleons available for interaction makes the process fully specified: after its creation from a reaction on the first nucleon, the pre-hadron can only interact with the other nucleon, see Fig.~\ref{fig:Frank1}. 
Another advantage of the deuteron is its small nuclear size, which can allow the longitudinal characteristic length of the vector meson production to exceed the nuclear size at intermediate energies, making the study more tractable at Jefferson Lab than for heavy nuclei.
The existence of the intact deuteron in the final state strongly limits the possibility of FSI, which minimizes theoretical uncertainties.
Thus, the deuteron is an ideal system for hadronization and CT studies {\it via} double-scattering (Fig.~\ref{fig:Frank1}{\color{blue}b}).
The magnitude and $t$-dependence of the  cross section provides information about the intermediate hadronic state and its evolution from a compact, color-singlet quark-gluon wave packet to a hadron.
Use of the deuteron as a target is already planned in the approved Hall C experiment E12-23-010 to study small size configurations of the proton using CT~\cite{Li:2023adv}.
However, a CT signal can be mimicked by another phenomenon, nuclear shadowing~\cite{Feynman:1973xc, Kopeliovich:2012kw}. Fortunately, as will be explained below, the latter can also be isolated and characterized with a tensor-polarized deuteron system~\cite{Frankfurt:1997ss}.
The phenomenon has the same underpinning as the nuclear shadowing first observed at low Bjorken-$x$ in the EMC deep-inelastic scattering (DIS) data~\cite{Frankfurt:1988nt, Norton:2003cb, Malace:2014uea} except that it involves hadronic d.o.f. rather than partonic ones.
The width of the wavefuntion of the probed nucleon (or parton in EMC's DIS case), which affects the spacetime resolution of the probe, is set by the characteristic longitudinal interaction length of the process, $\delta_z \simeq 2\nu/(m_V^2 \mp t)$ (or $\nu/Q^2$ in the DIS case), where $m_V$ is the mass of the vector meson and $-$ is for single scattering (Fig.~\ref{fig:Frank1}{\color{blue}a})
and $+$ for double scattering (Fig.~\ref{fig:Frank1}{\color{blue}b}), in the case of coherent meson production. As $\delta_z$ becomes comparable to the nuclear (or nucleon in the DIS case) size, the nucleons (partons) neighboring the struck nucleon (parton) also contribute to the reaction. The overlap of the nucleon (parton) wavefunctions results in a destructive interference~\cite{Feynman:1973xc} that reduces the cross section causing the so-called {\it shadowing}.
Clearly, this process is not a FSI, since it occurs in the initial step of the reaction. Therefore, shadowing is not considered a part of CT. Yet, since CT and shadowing effects may mimic each other, it is important to isolate the two effects in order to  unambiguously study CT or/and shadowing. $\rho$-meson production off a tensor-polarized deuteron target offers such a possibility by allowing the evolution of $\delta_z$ to be measured, since this is essentially the propagator entering into the scattering amplitudes of the processes shown in Fig.~\ref{fig:Frank1}{\color{blue}a} and \ref{fig:Frank1}{\color{blue}b}~\cite{Frankfurt:1997ss}. 
This, together with our good theoretical knowledge of the deuteron structure and the simplicity of that nucleus allows for a precise study of nuclear shadowing of hadronic d.o.f. 
In all, both nuclear shadowing with hadronic d.o.f. and double-scattering events off the deuteron, {\it i.e.}, the simplest occurrence of nuclear FSI and signal for CT, can be uniquely isolated and studied using intermediate energy ($\sim$10~GeV) real photons and a tensor-polarized deuteron target. 
\begin{figure}[htb]
\center
\includegraphics[width=0.75\textwidth]{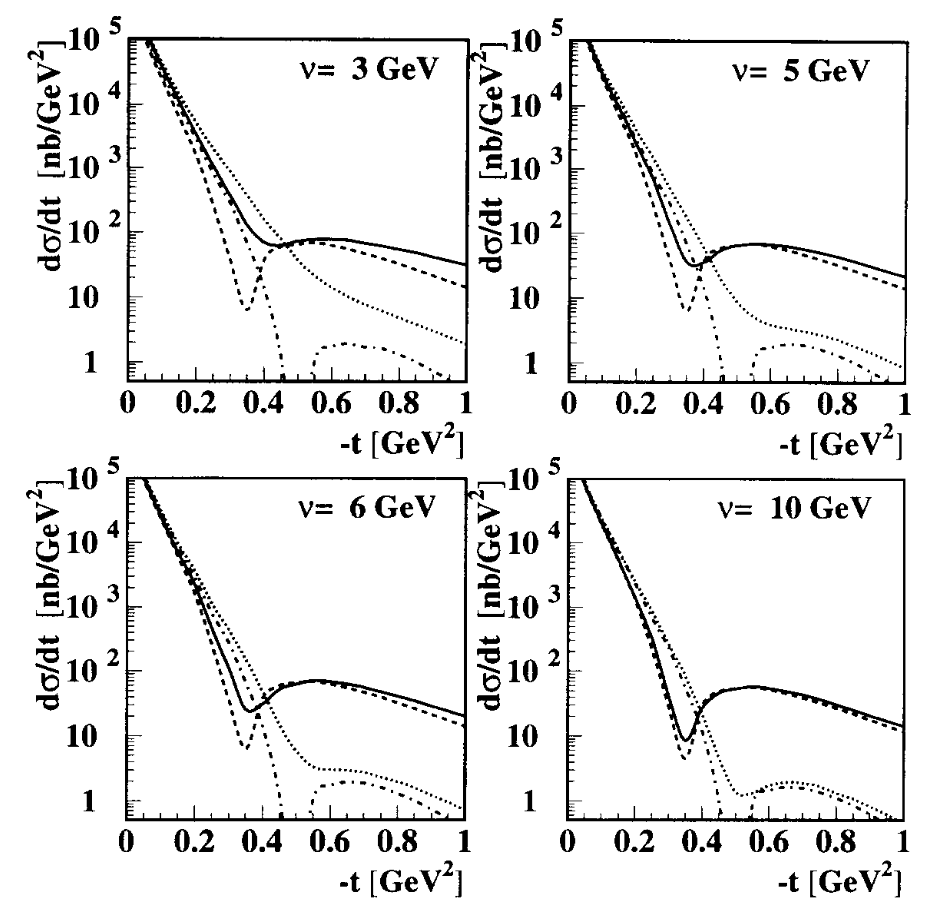}
\caption{\small Cross section $d\sigma/dt$ for $\gamma +\vec{d}\to \rho + d$ as a function of momentum transfer $-t$, with the deuteron initially in the $m=0$ state 
The four panels correspond to four incident photon energies $\nu$. 
Solid line: complete vector meson dominance calculation, 
dotted line: single scattering contribution, 
dot-dashed curves: single scattering calculations for infinite characteristic longitudinal interaction length $\delta_z$, dashed line: full calculations for infinite $\delta_z$. The cross section dip originates from the charge form factor node. Its depth depends on $\delta_z$ while its location reveals the double scattering process, \ref{fig:Frank7}{\color{blue}b}. 
(Figure from Ref.~\cite{Frankfurt:1997ss}.)}
\label{fig:Frank7}
\end{figure}
Fig.~\ref{fig:Frank7} displays the expected cross section 
for $\rho$ photoproduction, the easiest experimental case due to its large cross section and clear signature. The distinct effects of shadowing (viz $\delta_z$) and CT (viz FSI) are clear: a finite $\delta_z$ suppresses the dip depth (compare the dot-dashed curve with the dotted curve for a given $\nu$) and the FSI shifts its location (compare the dot-dashed curve with the dashed curve). Hence, measuring the evolution of the dip allows for a separation and determination of both nuclear shadowing and CT.

\section{Experiment\label{sec:Experiment}}

\subsection{Required Equipment}

\subsubsection{GlueX Spectrometer}

We plan to use the GlueX spectrometer, likely unchanged from the 2025 configuration (GlueX-II/JEF/Alpha) except for the addition of a polarized target.  The GlueX-I apparatus, described in detail in Ref~\cite{GlueX:2020idb}, has been modified by a substantial upgrade of the Forward Calorimeter.

\subsubsection{Beam} 

Circularly polarized photons are produced automatically from a standard radiator if the incident electrons are polarized. 
Electron beam polarization is not required in order to measure the $\phi-N$ cross section but is required in order to measure the helicity asymmetry $\mathbb{E}$ (Appendix~\ref{sec:add_physics:vectorpol}), which is an important component of the additional physics which can come out of this measurement.

We will run with the Hall D diamond radiator in order to produce linearly polarized photons in addition to circularly polarized ones.  The coherent peak containing the linearly polarized photons will be set at $\approx9$~GeV, which will maximize the flux by matching the beam spectrum to the Hall D tagger capabilities, and will allow measurement of the $\mathbb{G}$ asymmetry (Appendix~\ref{sec:add_physics:vectorpol}).
This will produce ``Elliptical" (both circular and linear) polarization at this energy~\cite{Afzal:2024dbo}.

\subsubsection{Target \label{sec:target}}

The target for this experiment will be that already approved for REGGE (E12-20-011~\cite{Dalton:2020wdv}).  
Reference~\cite{Dalton:2025jzt} describes the implementation of the target using the Hall D solenoid to provide most of the field needed to polarize the target. 

In this system, the target sample will be cooled to a temperature of about 0.2~K using a $^3$He--$^4$He dilution refrigerator inside the Hall D solenoid.
Thin, superconducting shim coils will be included within the cryostat to
increase the field around the sample to 2.5~T and improve its uniformity to the level required for dynamic nuclear polarization.  Microwaves near the electron resonance frequency of 70~GHz will be continuously applied to the sample, and based on the results of Goertz {\it et al.} \cite{GOERTZ200443}, we anticipate that vector polarizations
of $\pm$0.75 can be achieved at these field and temperature conditions.  According to Eq.~(\ref{eq:PzzfromPz}), the corresponding tensor polarization will in both cases be $Q=+0.48$.  These are the experimental conditions $Q_1$ and $Q_3$ indicated in Fig.~\ref{fig:triangle}.  

With the microwaves switched off, the target's dilution refrigerator will also be capable of cooling the samples to ultralow, millikelvin temperatures and operating in the so-called ``frozen-spin'' state.  The refrigerator constructed for the Hall B frozen spin target achieved an in-beam temperature of 32~mK~\cite{Keith:2012ad}.  At such low temperatures we can employ Adiabatic Fast Passage (AFP) to strategic portions of the NMR line and generate a negative tensor polarization up to -0.71, starting from either positive or negative vector polarization.  Following AFP the vector polarization will be reduced to about $\pm$0.15.  These are the conditions $Q_2$ and $Q_4$ in Fig.~\ref{fig:triangle}.   A tensor asymmetry between $Q_{1}$ and $Q_{2}$ could potentially include a substantial vector component.  A grand asymmetry including all 4 states will cancel any vector asymmetry.

Unlike $Q_1$ and $Q_3$, these conditions cannot be maintained continuously.  Instead both the vector and tensor polarizations will relax towards the very low equilibrium values (a few percent) determined by the field and temperature.  This relaxation is not well known when the deuteron level populations fall outside the ``Boltzmann curve'' of Fig.~\ref{fig:triangle}, but  it is expected to be highly temperature-dependent.  The available information does not extend to temperatures below 0.1~K and extrapolations to lower temperatures differ by an order of magnitude.  For this proposal we use the most conservative value of 10 hours.  
An R\&D program is underway at the University of Tennessee-Knoxville to optimize the AFP technique for generating negative tensor polarization in deuterons and study its relaxation at ultralow temperatures~\cite{NadiaFomin}.   

\begin{figure}[htb]
\center
\includegraphics[width=0.6\textwidth]{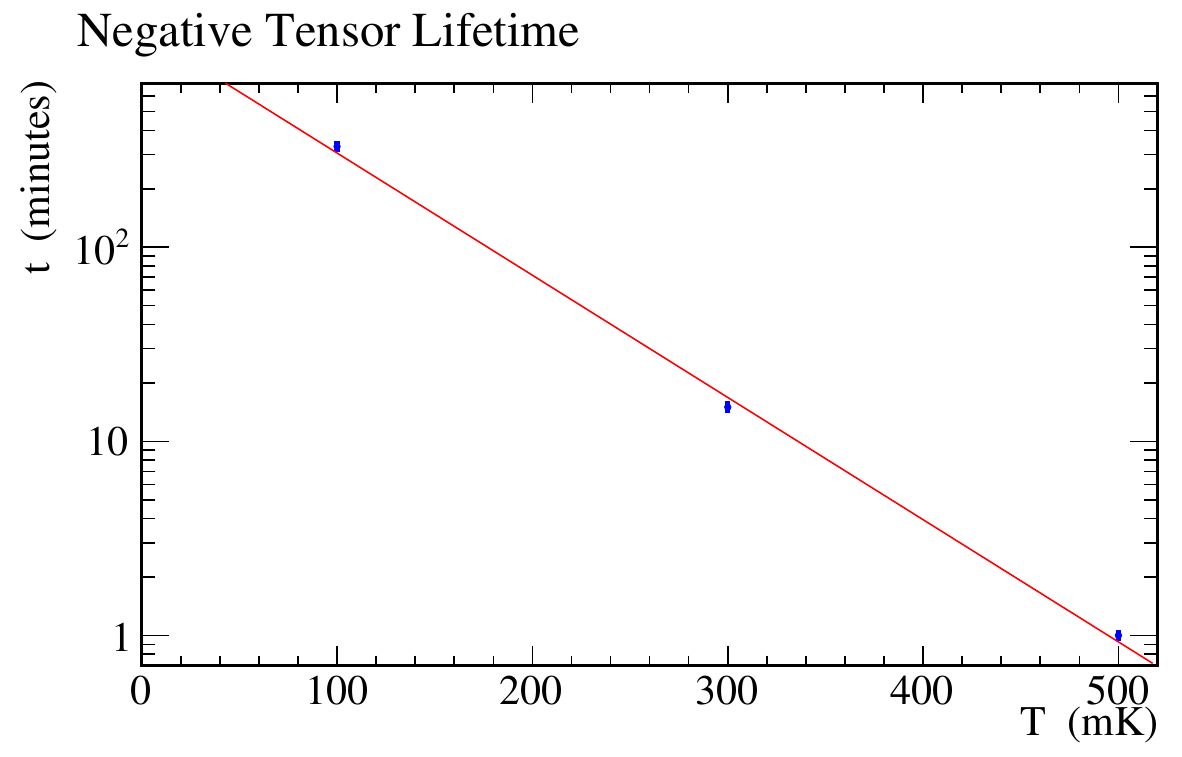}
\caption{\small Lifetime measurements from De Boer~\cite{deBoer:1973ft} are used to extrapolate to lower temperatures using a exponential model.}
\label{fig:extrap}
\end{figure}

The lifetime of the negative tensor polarization state has not been measured at the target temperature proposed here.
De Boer~\cite{deBoer:1973ft} showed a lifetime of 1~min at 0.5~K, 15~min at 0.3~K, and 330~min at 0.1~K.  In Fig.~\ref{fig:extrap}, we extrapolate these data for a conservative temperature of 50~mK, assuming constant relative uncertainty on each time measurement. 
This provides a lifetime of 10.5~hours.  

In calculating the target performance, we make the following conservative assumptions based on previous experience:

\begin{itemize}
    \item An asymptotic vector polarization of $P=0.75$ can be achieved, which corresponds to a tensor polarization of $Q=0.48$.  The FROST target in Hall B~\cite{Keith:2012ad} achieved a value of $P=0.87$ at higher field (5~T) but higher temperature (0.3~K), and \cite{Goertz:2004} achieved $P=0.8$ at the 2.5~T and 0.15~K, which will be the setting of the Hall D target.  
    \item The positive tensor state (positive or negative vector polarization state) will be continuously pumped with a spin-up time-constant for producing vector polarization of 4 hours, projecting from Ref.~\cite{Goertz:2004}. 
    \item A model of AFP, described in Ref.~\cite{EPJA}, is used to determine how much negative tensor polarization will be achieved from a given vector polarization.  For example, from $P=0.75, Q=0.48$. (A perfect manipulation would give $Q=-0.71$.)
    \item An efficiency of 90\% is assumed when manipulating spins using AFP to go from positive tensor to negative tensor or vice versa. (91\% was achieved~\cite{Hautle:1992} going from positive vector to negative vector, which is a more complicated manipulation.)
    \item One hour before going from positive to negative vector polarization, the pumping microwaves will be turned off to allow the target to cool to 50 mK, which will interrupt the polarization growth.
    \item The beam will remain continuously on the target, and we will average over the evolving polarization.
\end{itemize}

\begin{figure}[htb]
\center
\includegraphics[width=0.6\textwidth]{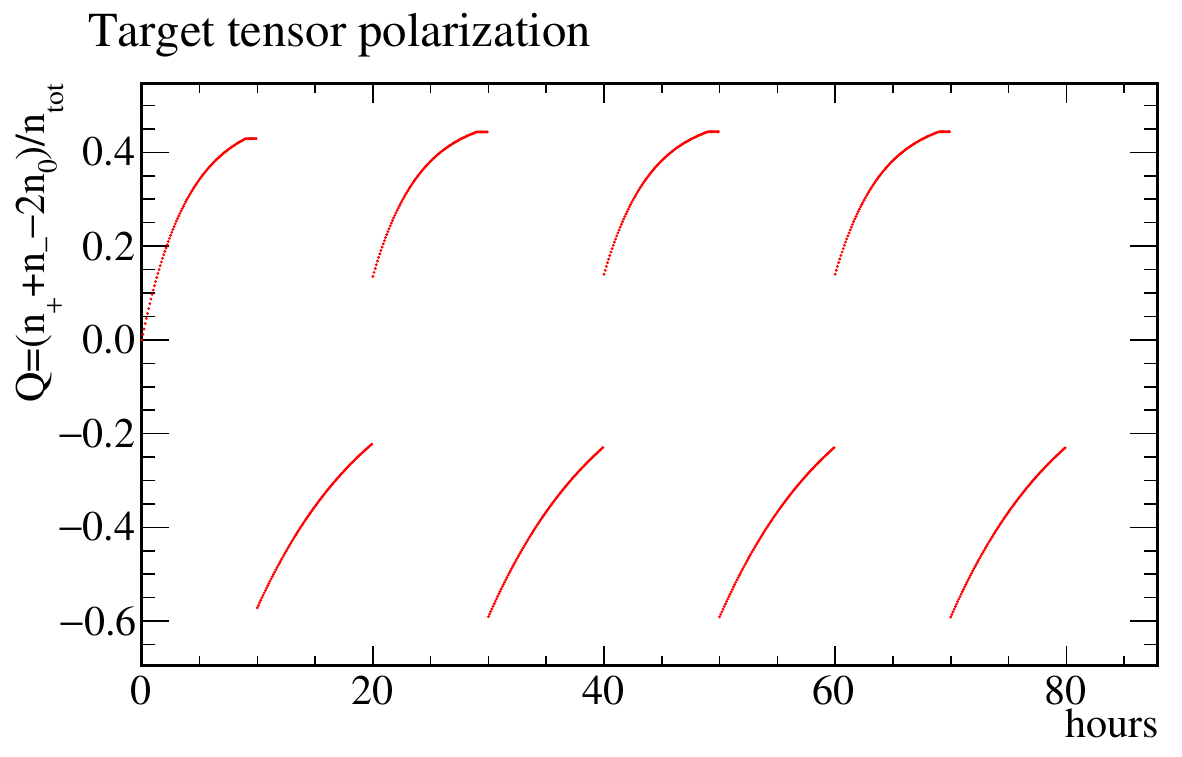}
\caption{\small Tensor polarization of the target versus time under the running conditions described in the text.  Shown are 4 of the 20-hour cycles.  Starting from zero polarization, the steady state is achieved after 1 cycle.}
\label{fig:target_sequence}
\end{figure}

Given these assumptions, the experiment is optimized with the same run-time in each of the four states, $Q_1$ through $Q_4$.  This allows the use of the most straightforward asymmetry formulae.  Running longer will allow more pumping in the  $Q_{1,3}$ states and will therefore increase the initial polarization for the $Q_{2,4}$ states.  It would also allow more decay during the $Q_{2,4}$ states and therefore also lowers the initial polarization of the $Q_{1,3}$ states.  This strong coupling between the states leads to an optimum cycle length of about 20 hours with average values for the polarizations of $\bar{Q}_{1,3}=+0.35$ and $\bar{Q}_{2,4}=-0.38$.  These values are used to determine the expected results in Sec.~\ref{sec:results}.  Fig.~\ref{fig:target_sequence} shows $Q$ versus time for this scenario.

\FloatBarrier

\subsection{Statistical Uncertainties}

The expected number of events is obtained by scaling from existing Hall D data taken on a liquid deuterium target. We will use the same standard equipment (the Forward Calorimeter was upgraded in 2023-2025 but this will not especially benefit our reactions) and thus, our uncertainty estimate is as accurate as it is possible.  This is described in detail in Sec.~\ref{sec:projection}.

\FloatBarrier
\subsection{Systematic Uncertainties \label{sec:Systematics}}

\subsubsection{Beam Flux}

In an asymmetry measurement, it is necessary to normalize the data in each period by the beam flux. The latter represents an
important systematic uncertainty for tensor-polarized targets used with electron beams~\cite{PR12-13-011_Slifer,PR12-15-005_Long}.  Since the target remains in a given polarization state for hours to days, the flux normalization is performed over these timescales, which introduces a sensitivity to drifts in the flux monitor.
In Hall D, the photon beam flux is monitored primarily using a Pair Spectrometer by directly sampling the photon beam and counting individual $e^+e^-$ pairs from individual photons~\cite{Barbosa:2015bga}.  As a counting measurement with large signals, this method has negligible baseline drifts or threshold effects, and the counting statistics of normalization events are many orders of magnitude greater than the physics events.
Thus, the large asymmetries of this measurement along with the counting measurement technique renders the flux uncertainty negligible.
For photon energies below the acceptance of the Pair Spectrometer, we will use the cross section for Compton scattering ($\gamma e^- \to \gamma e^-$) measured in the GlueX spectrometer concurrently with the reactions of interest.
The Compton cross section has been shown to be a good luminosity monitor by the Hall D PrimEx experiment~\cite{Compton}.
Only a relative normalization is required for this purpose.
We include 2 PAC days of dedicated time to change the PS acceptance and study Compton events and the PS response simultaneously. This will allow to reach the 2\% uncertainty already achieved in the PrimEx experiment~\cite{Compton}.

\subsubsection{Target Polarization}

The standard technique for measuring vector polarization in a polarized target uses the integrated area under the NMR line, and this must be calibrated against a known polarization, usually the thermal equilibrium polarization at 1~K.  This calibration can be determined with a relative uncertainty of $dP/P\approx4\%$~\cite{Keith:2003ca}. 
For the positive tensor polarization state, where $Q^+$ is determined from Eq.~(\ref{eq:PzzfromPz}), the uncertainty scales as $dQ/Q^+\approx2 dP/P$ so we assume $dQ/Q^+\approx8\%$.
For the unpolarized state, the residual vector and tensor polarizations are both less than 0.01, and their uncertainties are negligible. 
For the negative tensor polarization state, the area-based technique will not work and $Q$ must be determined from a NMR line shape analysis.
Reference~\cite{SpinMuon:1997jxq} showed agreement between the line shape and TE-calibration methods at the 3\% level and, more generally, suggests that the line shape method can probably achieve 5\% or better.
Here, we assume that it can be made equal in precision and that it is fully anti-correlated, so $dQ/Q^-\approx8\%$.  This leads to a normalization uncertainty of $dA_{zz}/A_{zz}\approx8\%$ from both $Q_1=0.48, Q_2=0$ and for $Q_1=0.35, Q_2=-0.38$.  

In general, the observable of interest in coherent vector meson production is sensitive to the shape of the asymmetry differential in $-t$, and has little sensitivity to the overall normalization.  
The target polarization uncertainty does not, to first order, affect the location of minima, maxima or zero crossings in $-t$ and so will not be a leading uncertainty in the extraction of physical observables from the data.
The exception is for the $\rho$ measurement described in Section~\ref{sec:physics:rho}.
Although the primary interest is the change in the size of the $A_{zz}$ peak with energy, a determination of the absolute peak size in $A_{zz}$ is also valuable. 

\subsubsection{Beam Polarization}

The beam polarization uncertainty enters in measurements of the double-polarization observables $\mathbb{E}$ and $\mathbb{G}$.  Since there will already be a $4\%$ uncertainty on the target vector polarization the requirements on the knowledge of the beam polarization are modest.
We request a bi-weekly measurement of the electron beam polarization using the Mott polarimeter in the injector.  This could be avoided if any other halls measure the polarization. 
The photon beam circular polarization in Hall D can be obtained from a combined fit of polarization measurements in other halls~\cite{Grames:2004mk}.  This leverages the well-understood systematic uncertainties of the M\o ller polarimeters in other halls to achieve a polarization systematic uncertainty in Hall D of $dP_\odot/P_\odot<2\%$ depending on the orthogonality of the fit.

The linear polarization of the photon beam is measured unobtrusively using the Triplet Polarimeter (TPol) which uses the process of $e^+e^-$ pair production on atomic electrons in a beryllium target foil.
The estimated total systematic uncertainty is 1.5\%~\cite{Dugger:2017zoq} and over the course of a run period of a few months, a total uncertainty of 2.1\% has been achieved~\cite{GlueX:2017zoo}.

\subsubsection{Backgrounds}
\label{background-estimate}
Thanks to kinematic constraints from the exclusivity and the ability to unambiguously tag a recoil deuteron using $dE/dx$ in the Central Drift Chamber, coherent production from the deuteron has been observed in existing Hall D data to have very little background, for both $\phi$ and $\rho$ production (see Section~\ref{sec:projection}).

\subsubsection{Total systematic uncertainty}

\begin{table}[htb!]
    \caption{Table of systematic uncertainties for the measurements.}
    \label{tab:systematics}
\centering
\begin{tabular}{lcc}
& Quantity     & Uncertainty (\%)     \\
\hline
Target polarization        &   $dQ/Q^-$           &   8          \\
Charge normalization       &   $dN/N$             &   2          \\  
Total                      &   $dA_{zz}/A_{zz}$   &   8     \\
\end{tabular}
\end{table}

The total systematic uncertainty on the $dA_{zz}/A_{zz}$ is an overall normalization uncertainty.  As will be seen in Sec.~\ref{sec:results:phi}, the observable of interest is the location of a feature versus $-t$ and which is insensitive to an overall normalization uncertainty.

\FloatBarrier

\section{Expected Results\label{sec:results}}

\subsection{Projections from existing data}
\label{sec:projection}

To project the results of our proposed measurement, we use Hall D data from the 2021  run on an unpolarized liquid deuterium target, currently under analysis by the SRC-CT Collaboration.  
During that time, 2.9 PAC days of deuteron data were taken.
The kinematics and target nucleus are the same as for our proposal, providing us with the opportunity to use the most accurate method to estimate the rates.

Our rates are predicted by looking at real data for the final states, $\gamma D\to \phi D \to K^+K^- D$ and $\gamma D\to \rho^0 D \to \pi^+\pi^- D$.  
No additional tracks or showers are permitted.  
These are fully exclusive final states which allow for a complete, 4-component kinematic fit.
This is required to have a p-value of at least 0.01. 
A cut on missing energy is applied as well as
a PID cut on the $K$ or $\pi$ mesons requiring that the time of flight and $dE/dx$ in the chambers be consistent with the correct particle with a p-value of at least 0.01, given the known resolutions.

\begin{figure}[htb]
\center
\includegraphics[width=0.99\textwidth]{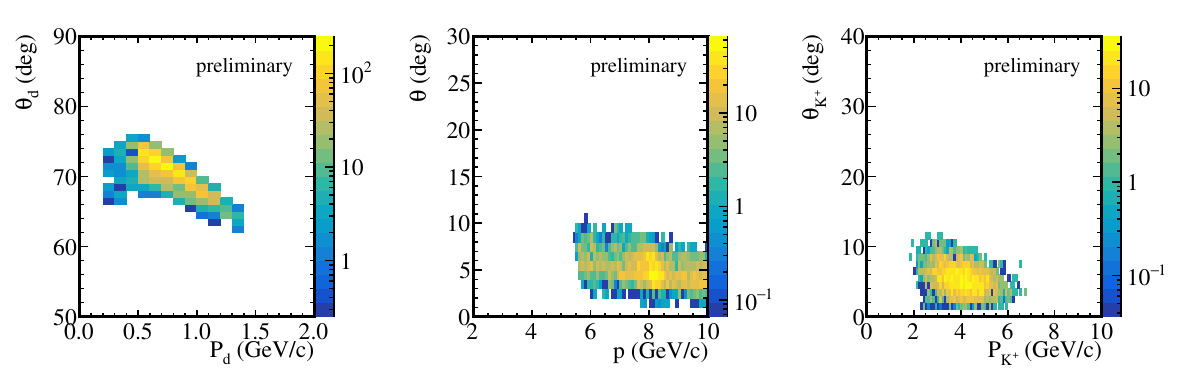}
\caption{\small The kinematics of the particles in the reaction $\gamma D\to \phi D \to K^+K^- D$, for the $D$, $\phi$ and $K^+$ respectively.  Preliminary data from the 2021 SRC-CT experiment.}
\label{fig:SRC_phi2H_kin}
\end{figure}
\begin{figure}[htb]
\center
\includegraphics[width=0.99\textwidth]{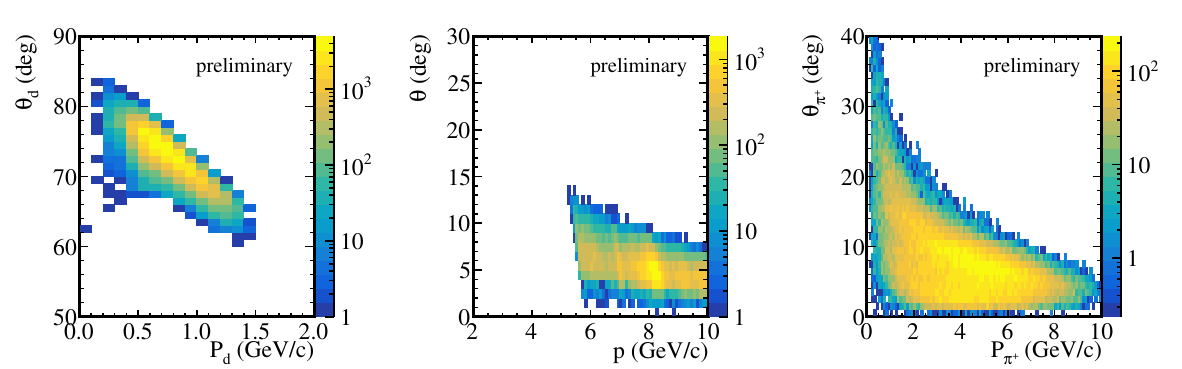}
\caption{\small  The kinematics of the particles in the reaction $\gamma D\to \rho^0 D \to \pi^+\pi^- D$, for the $D$, $\rho$ and $\pi^+$ respectively.  Preliminary data from the 2021 SRC-CT experiment.}
\label{fig:SRC_rho2H_kin}
\end{figure}
\begin{figure}[htb]
\center
\includegraphics[width=0.8\textwidth]{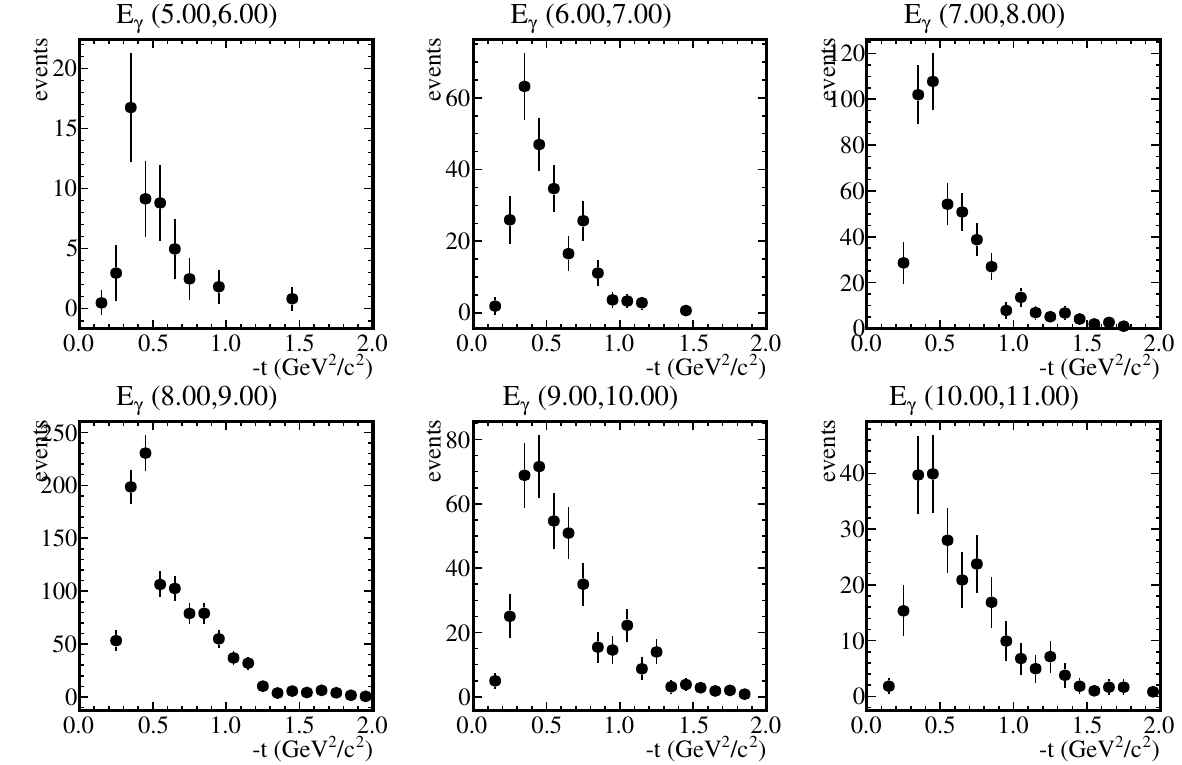}
\caption{\small  Total number of coherent $\gamma d \to \phi d$ events detected for 2.9 PAC days of beam time during the 2021 SRC-CT experiment run on unpolarized liquid deuterium (preliminary data).  These data predict the number events expected from this proposal.
}
\label{fig:phi_events_SRC}
\end{figure}
\begin{figure}[htb]
\center
\includegraphics[width=0.8\textwidth]{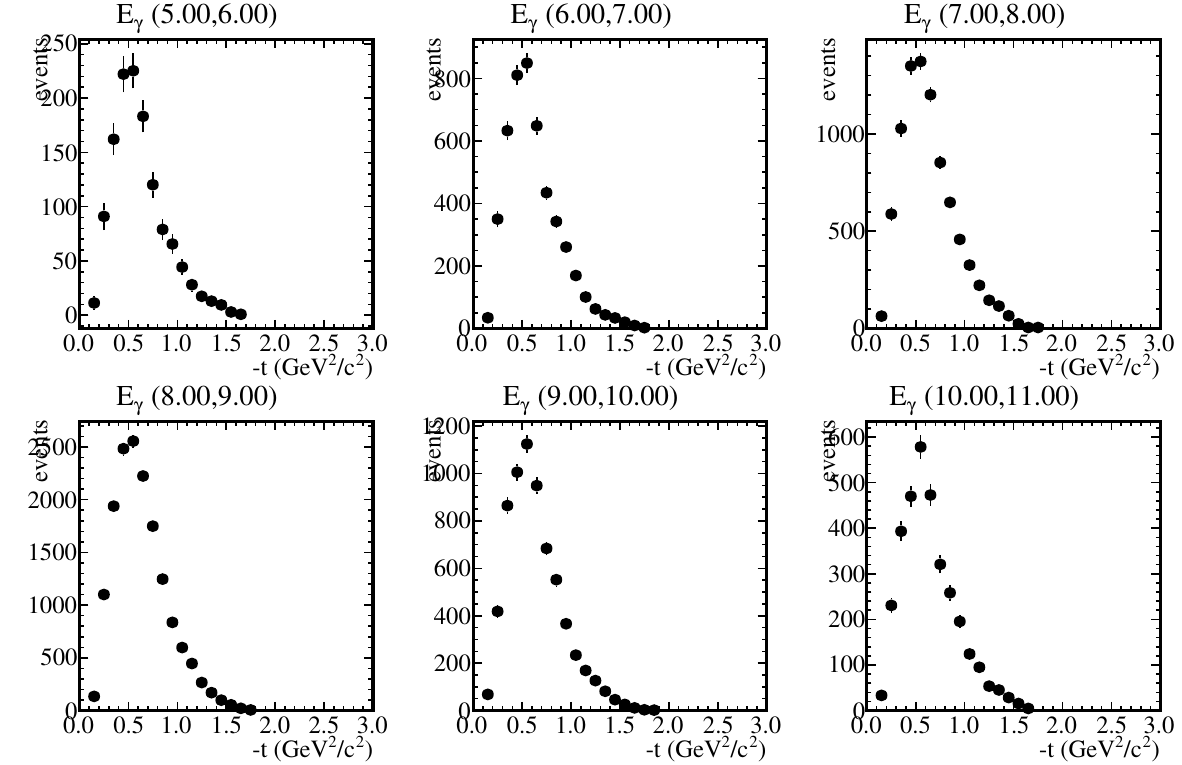}
\caption{\small Same as Fig.~\ref{fig:phi_events_SRC} but for $\gamma d \to \rho d$.
}
\label{fig:rho_events_SRC}
\end{figure}

The kinematics for the accepted events in those  reactions are shown in Figs.~\ref{fig:SRC_phi2H_kin} and \ref{fig:SRC_rho2H_kin}.
The deuteron has limited acceptance until it reaches a momentum of 400 MeV.  
The deuteron and vector meson are very closely correlated, consistent with a 2 body final state.
The observed wide momentum range is due to the large range of beam energies.
\begin{figure}[htb]
\center
\includegraphics[width=0.8\textwidth]{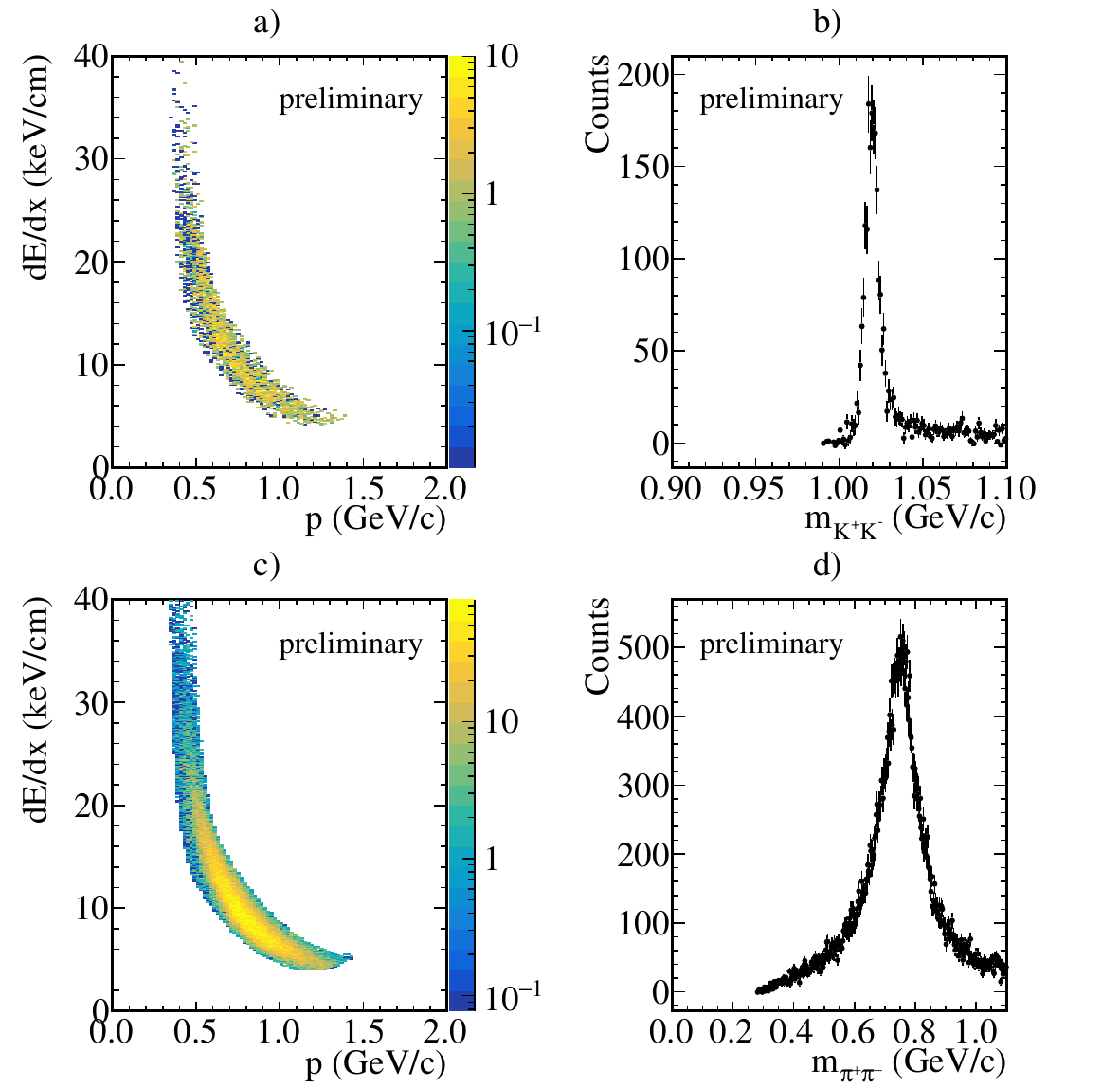}
\caption{\small Left: $dE/dx$ vs deuteron momentum $p$ in the Central Drift Chamber, right: reconstructed mass of vector meson. Top row: $\gamma D\to \phi D \to K^+K^- D$, bottom row: $\gamma D\to \rho^0 D \to \pi^+\pi^- D$.
Preliminary data from the 2021 SRC-CT experiment.}
\label{fig:SRC_gen2H}
\end{figure}
A very important cut is that on $dE/dx$ in the Central Drift Chambers (CDC), shown in Fig.~\ref{fig:SRC_gen2H} in panels a) and c).  
The deuteron has higher ionizing power and lower momentum than a proton, allowing such a clean  separation by the $dE/dx$ cut that non-deuteron background is negligible.
Since the deuteron does not have a nuclear excited state that is bound, the presence of an intact deuteron shows that it is in the ground state.  
Furthermore, familiar sources of background for a proton target are absent here because any
nucleonic excitation (baryon resonances) would destroy the deuteron.
The accidental interchange of the positively charged tracks (a $\pi^+$ or $K^+$ with a deuteron) is also not possible because the deuteron is so clearly identified as such.
Thus, the data has very little background, as seen in Fig.~\ref{fig:SRC_gen2H} b) and d).
The observed event distributions for $\gamma d \to \phi d$ in bins of beam energy and $-t$ is shown in Fig.~\ref{fig:phi_events_SRC}, and for $\gamma d \to \rho d$ in Fig.~\ref{fig:rho_events_SRC}.

We smooth the above distributions to suppress statistical fluctuations and then scale them to account for the different experimental configurations between SRC-CT and this proposal. 
The 10 cm long polarized target is composed primarily of butanol and will contains 1.7 times the number of nucleons compared to the liquid target. However, only 25\% are deuterons, 
which leads to to an areal density that is a factor of 0.4 lower than that of the liquid target.
Overall, the hadronic rate will be larger by a factor of 1.7, requiring us to reduce the photon flux.
We plan to run at a luminosity of 1.3 times the SRC-CT luminosity by optimizing the trigger and recording at higher DAQ rate.
Given the SRC-CT 2.9 PAC days of D run, and that we will request 60 PAC days for production, 
this leads to an overall scaling of a factor of 6.3.

\FloatBarrier
\subsection{Coherent photoproduction of $\phi$}
\label{sec:results:phi}

\begin{figure}[htb]
\center
\includegraphics[width=0.80\textwidth]{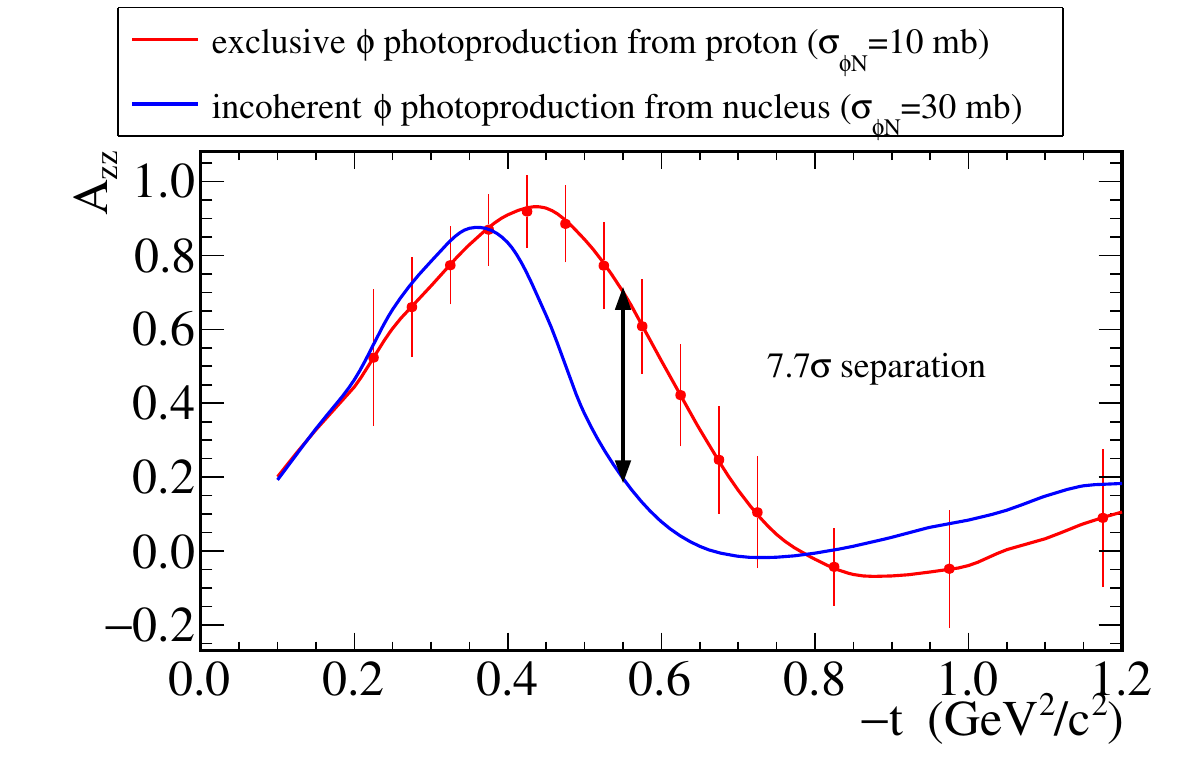}
\caption{\small The tensor asymmetry $A_{zz}$ for $\gamma d \to \phi d$ with 60 days of beam time on the 10~cm polarized deuteron target, combined in a single bin of beam energy between 5 and 11 GeV, to provide the full statistical power. 
The two curves are calculations~\cite{Frankfurt:1997ss} using either the value of $\sigma_{\phi N}$ determined from $\phi$ photoproduction off the proton (red curve) or off nuclei (blue curve), with both cases describing equally well the CLAS data~\cite{CLAS:2007xhu} of Fig.~\ref{fig:coherent_unpol}. Our expected data (red circles) have been placed arbitrarily on the $\sigma_{\phi N}=10$ mb scenario (red curve).
The uncertainties shown are statistical and have a discriminating power of 7.7$\sigma$ between the two scenarios. 
The systematic uncertainties are either fully point-to-point correlated (polarimetry, beam flux), rescaling the vertical axis and therefore irrelevant to the discriminating power  of this measurement, or are negligible (see Section~\ref{background-estimate}).  
}
\label{fig:Azz_phi_results}
\end{figure}

The number of detected events for $\gamma d \to \phi d$ from the 2.9 PAC days of SRC-CT is shown in Fig.~\ref{fig:phi_events_SRC}.  
\begin{figure}[htb]
\center
\includegraphics[width=0.8\textwidth]{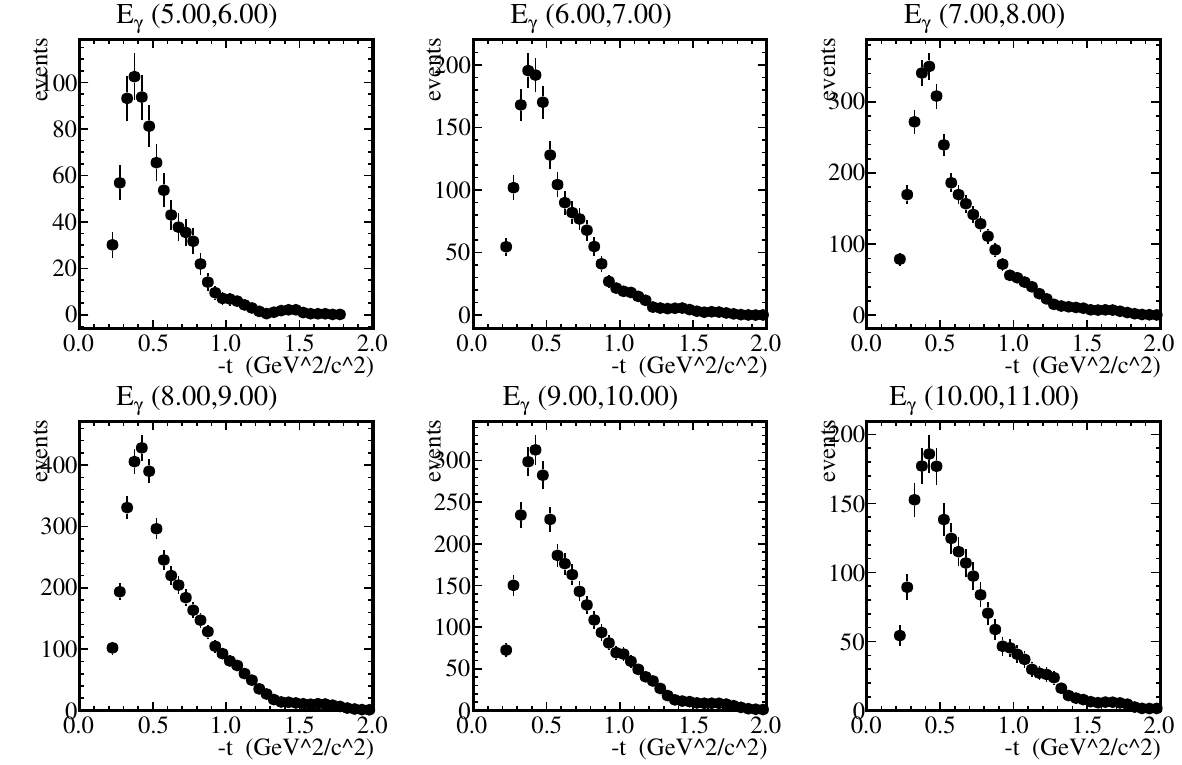}
\caption{\small Expected number of coherent $\gamma d \to \phi d$ events as a function of $-t$ in bins of beam energy for 60 days of beam time on a polarized deuteron target. }
\label{fig:phi_events_proj}
\end{figure}
%
From this we obtain the number of events for the same reaction expected for this experiment, see Fig.~\ref{fig:phi_events_proj}, which, together with Eq.~(\ref{eq:Azz_2qvals}) and the target parameters from Sec.~\ref{sec:target}, determine the statistical power to measure the tensor asymmetry $A_{zz}$ versus $-t$.  
This is shown in Fig.~\ref{fig:Azz_phi_results} for all data integrated over the beam energy.
We also show two curves from the model described in Sec.~\ref{sec:physics:tensor}, for the exclusive photoproduction from proton ($\sigma_{\phi N}=10$~mb and $b_{\phi N}=6$ (GeV/c)$^{-2}$), and for incoherent photoproduction from nuclei ($\sigma_{\phi N}=30$~mb and $b_{\phi N}=10$ (GeV/c)$^{-2}$).
These are the two scenarios considered in Ref.~\cite{CLAS:2007xhu} and they describe equally well the data in Fig.~\ref{fig:coherent_unpol} (left).
The power of the data to discriminate between the scenarios is 7.7$\sigma$.
A fit of the model to the data allowing only $\sigma_{\phi N}$ to vary would give a uncertainty of $\Delta\sigma_{\phi N}\approx1.5$\,mb for data centered on $\sigma_{\phi N}=10$~mb, and an uncertainty of $\Delta\sigma_{\phi N}\approx3.0$\,mb for data centered on $\sigma_{\phi N}=30$~mb.
A full fit would also need to vary $b_{\phi N}$.  While this may somewhat decrease the significance, it is still more than sufficient to discriminate between the two current available values of $\sigma_{\phi N}$, 10 mb and 30 mb.

\FloatBarrier
\subsection{Coherent photoproduction of $\rho$}
\label{sec:results:rho}

\begin{figure}[htb]
\center
\includegraphics[width=0.99\textwidth]{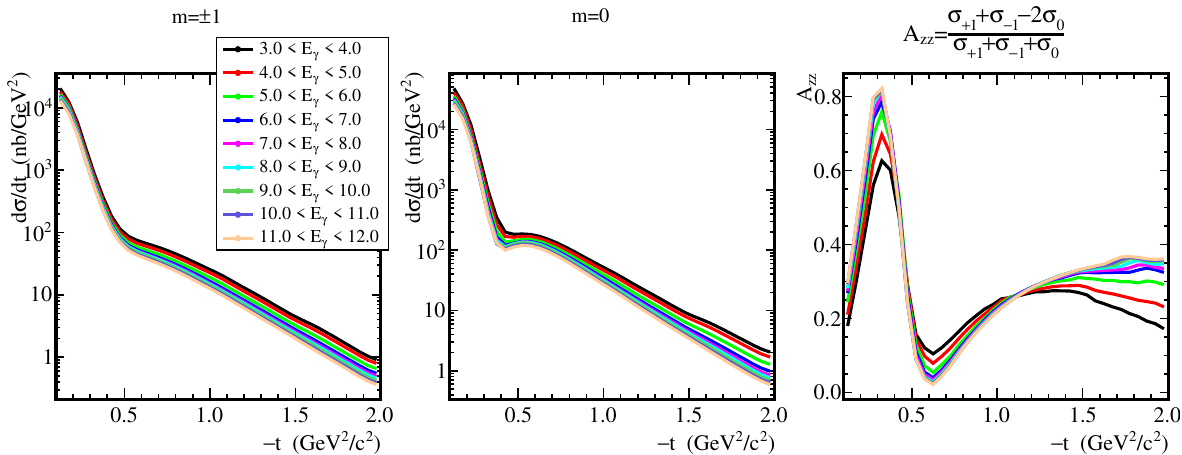}
\caption{\small Cross sections (left panel: $m=\pm1$ spin projections, center panel: $m=0$) and  tensor asymmetry (right panel) for $\gamma d \to \rho d$ as determined from Ref.~\cite{Frankfurt:1997ss}.
The evolution of the feature at $-t\approx0.3~\mathrm{(GeV/c)}^2$ with beam energy, from 3 GeV to 6 GeV, is due to the longitudinal coherence length of the photon.  
This effect can mimic color transparency in electroproduction~\cite{HERMES:1998ajz} and has to be considered in those measurements~\cite{CLAS:2012tlh}.
A quantitative test of our understanding of this effect in a very well controlled environment is possible in this experiment.
}
\label{fig:rho_Azz}
\end{figure}

We use the updated version of the calculation from Ref.~\cite{Frankfurt:1997ss}, see Section~\ref{sec:physics:phiN}, to compute the differential cross section over our energy range for the 3 spin states, and to determine the  tensor asymmetry from $A_{zz}=\frac{\sigma_{+1}+\sigma_{-1}-2\sigma_{0}}{\sigma_{+1}+\sigma_{-1}+\sigma_{0}}$, see Fig.~\ref{fig:rho_Azz}.  The dip feature in the $m=0$ state causes a large positive $A_{zz}$ at $-t\sim0.4$~GeV$^2$, which grows with increasing energy.

We used the SRC-CT 2021 data shown in Fig.~\ref{fig:rho_events_SRC}, scaled for beam and time as previously described, to obtain the expected number of $\gamma d \to \rho d$ events for 60 days of beam time on the 10~cm polarized deuteron target.
\begin{figure}[htb]
\center
\includegraphics[width=0.8\textwidth]{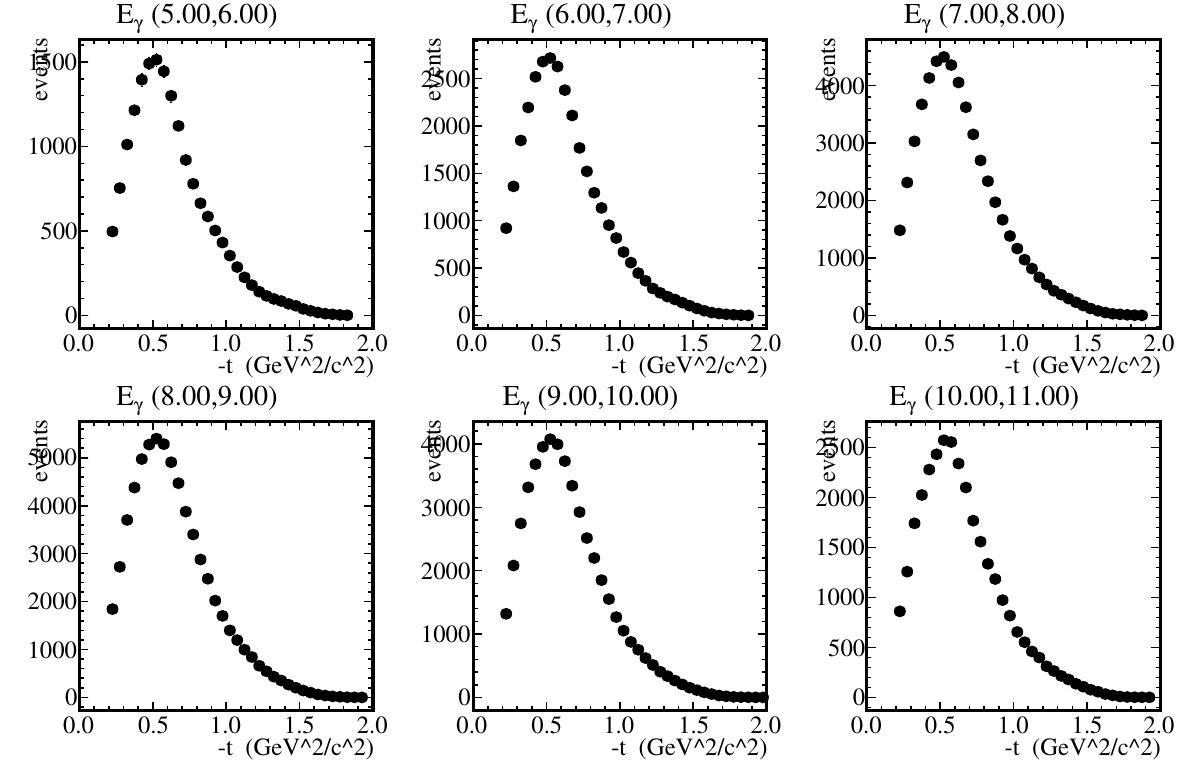}
\caption{\small Expected number of coherent $\gamma d \to \rho d$ events for 60 days of beam time on the 10~cm polarized deuteron target.
}
\label{fig:rho_events_proj}
\end{figure}

\begin{figure}[htb]
\center
\includegraphics[width=0.8\textwidth]{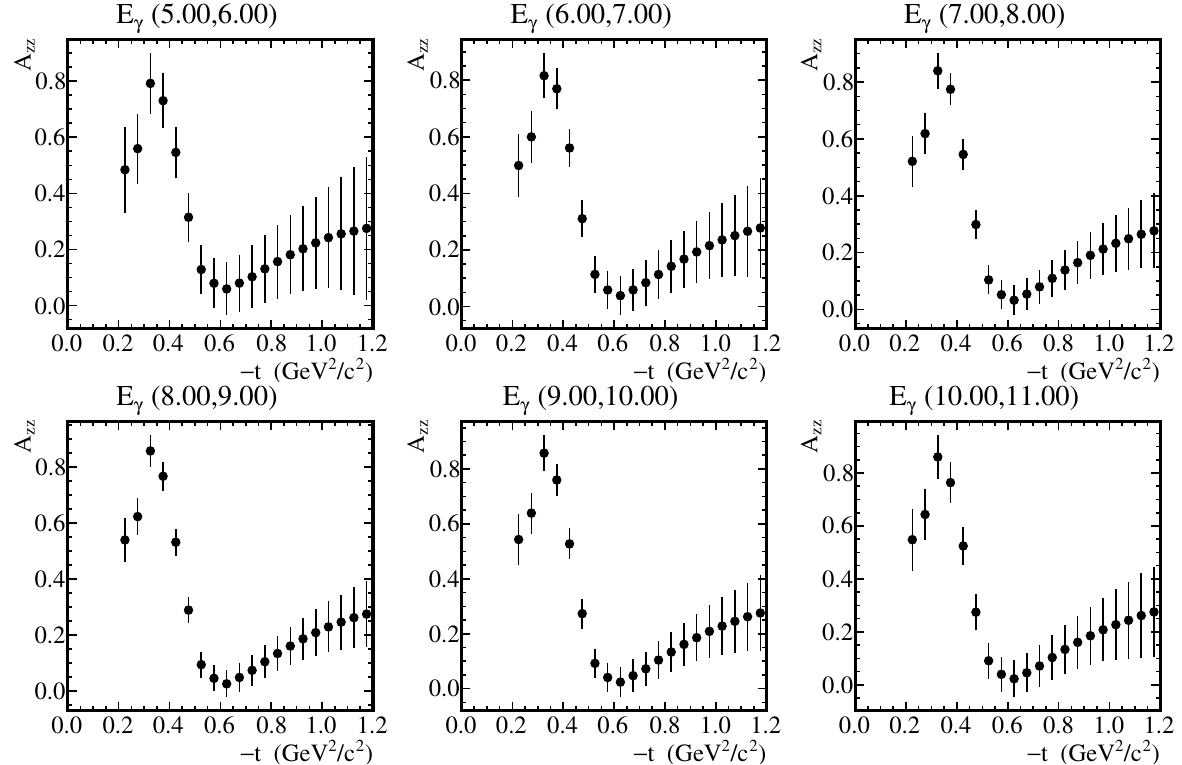}
\caption{\small The tensor asymmetry $A_{zz}$ as measured in $\gamma d \to \rho d$ with 60 days of beam time on the 10~cm polarized deuteron target for average tensor polarizations of $Q=0.35$ and $Q=-0.38$.
The dependence on $E_\gamma$ of the peak ($-t\simeq 0.3$~(GeV/c)$^{2}$) and the trough ($-t\simeq 0.6$~(GeV/c)$^{2}$) of $A_{zz}$  is expected to be even larger for the lowest energy bins covered by the Hall D tagger, viz 3--4~GeV and  4--5~GeV, see right panel of Fig.~\ref{fig:rho_Azz}. (Those expectations are not shown here since the comparable data were not taken by SRC/CT.) 
The precision of this data permits a careful test of any models that may used to extract information from the $\phi$ data.
}
\label{fig:asym_Q35Qm38_rho}
\end{figure}

$A_{zz}$ is displayed in Fig.~\ref{fig:asym_Q35Qm38_rho}, which shows how precisely we can map the tensor asymmetry versus both energy and $-t$.  We are very sensitive to effects that might shift the position or depth of the diffractive minimum in this process.

\FloatBarrier
\section{Beam Request\label{sec:Beam-time}} 

We request 60 days of beam to make measurements of multiple final states from tensor and vector polarized deuteron.  This number of days is needed to ensure a statistically significant separation of the two scenarios, a small $\sigma_{\phi N}=10$\,mb as indicated using vector meson dominance and exclusive phi photoproduction from the proton, and a large $\sigma_{\phi N}=30$\,mb as indicated from incoherent nuclear photoproduction.
Optimally, this experiment will be run in a suite of polarized target experiments in Hall D and thus, will not need dedicated target commissioning time.
We will require 1 day every 2 weeks to calibrate the NMR polarimetry.  This will be done during accelerator maintenance days, to save time.  We will using the Total Absorption Counter (TAC) as required to calibrate the Pair Spectrometer so that absolute cross sections can be extracted.  Table~\ref{tab:beam_time} summarizes this request.

Electron beam polarization is required to measure the double spin asymmetry $\mathbb{E}$ and  test $s$-channel helicity conservation.  The diamond radiator is required to optimize the flux to the tagger capabilities and to measure the double spin asymmetry $\mathbb{G}$.

\begin{table}[htb!]
    \centering
\begin{tabular}{cc}
Activity  & Beamtime   \\
\hline
Production data taking              & 60    \\
Target NMR calibrations             & 2     \\
Luminosity studies                  & 3     \\
\hline
Total Time                          & 65    \\
\end{tabular}
    \caption{Beam time request for the tensor-polarized target experiment.}
    \label{tab:beam_time}
\end{table}

\section{Summary \label{sec:Summary}}

We have presented a proposal that will finally resolve the longstanding puzzle of the $\phi$-nucleon cross section $\sigma_{\phi N}$ and unambiguously measure it to high precision. 
Additionally, the abundant data from coherent $\rho$ photoproduction will enable detailed studies of the coherence effects of the photon.
Our proposed experiment exemplifies the well-known observation that polarization offers additional leverage that further constrain or test theories -- in our case, specifically addressing the problem of determining $\sigma_{\phi N}$.

The experiment will use the polarized target being constructed for Hall D~\cite{Dalton:2020wdv}.  For negative tensor polarization it will be employed in frozen spin mode, which does not present particular difficulty and has been demonstrated in Hall B. 
Negative tensor polarization has been demonstrated before, although the lifetime of a negative tensor-polarized state for our lower temperatures has not been measured.  We use a conservative estimate and show that it will allow the goals of the experiment to be achieved.
Aside from the target, all other aspects of the experiment use standard equipment and have already been demonstrated, including detecting recoil deuterons down to sufficiently low momentum.
The GlueX spectrometer and beam energy range available in Hall D are ideal for the goals of the experiment, 
which will be the first to ever use nuclear tensor polarization with a real photon beam.

\appendix

\section{Additional Physics Opportunities \label{sec:add_physics}}

Here we briefly describe other physics that might be accessible with a tensor polarized target. We do not provide realistic models or projected uncertainties for these ideas since this proposal focuses on $\sigma_{\phi N}$ and coherent $\rho$ production.

\subsection{High Momentum Components in Deuteron Breakup or Non-Nucleonic Degrees of Freedom}
\label{hidden color}
\subsubsection{$\phi$ production in deuteron breakup}

A nucleus is often described as a collection of color-singlet nucleons.
The concept of hidden color configurations~\cite{Brodsky:1976mn, Matveev:1977xt, Harvey:1981udr} refers to the possibility of non-nucleonic d.o.f such as exotic nuclear states or nuclear structures that may involve color-exotic combinations of quarks, such as $\Delta$-isobars, six-quark states or hexadiquarks. Such states are expected in QCD~\cite{Gross:2022hyw, West:2020rlk}.
The tensor polarized deuteron is considered a promising place to look for signatures of hidden color~\cite{Miller:2013hla}.

\begin{figure}[htb]
\begin{subfigure}{.49\textwidth}
  \centering
  \includegraphics[width=.8\linewidth]{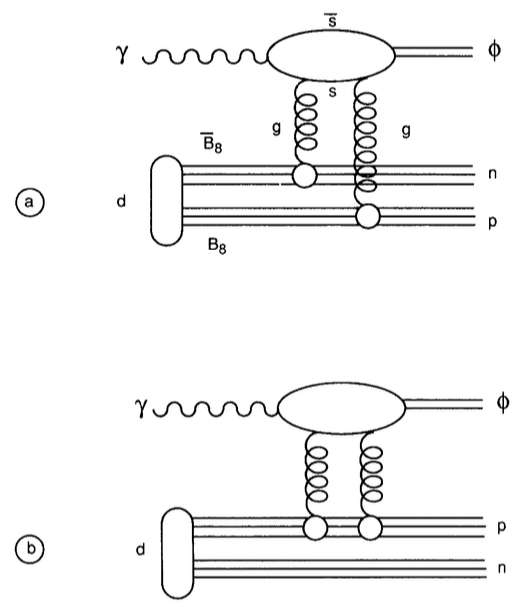}
\end{subfigure}%
\begin{subfigure}{.49\textwidth}
  \centering
  \includegraphics[width=.8\linewidth]{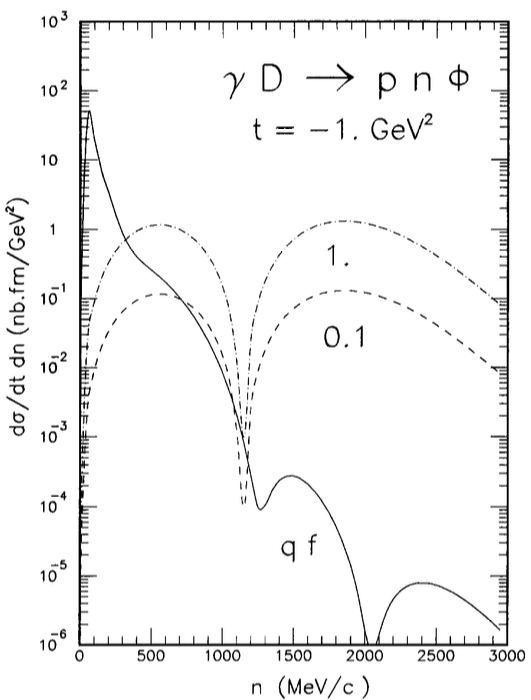}
\end{subfigure}
\caption{
Left: $\phi$ photoproduction from deuteron: a) two-gluon-exchange mechanism with each gluon coupled to a quark from a different
nucleon; b) standard, quasi-free, two-gluon-exchange mechanism.  
Right: $D(\gamma,p\phi)n$ reaction cross section vs the recoil-neutron momentum at $-t=1~\textrm{GeV}^2$. Solid line: quasi-free contribution. Dashed lines: contribution of the hidden color
component 0.1\% or 1\%. Figures from Ref.~\cite{Laget:1994ba}.
}
\label{fig:Laget_hiddencolor}
\end{figure}

An early idea to search for non-nucleonic d.o.f is to look for unexpected strength at kinematics where two nucleons have high relative momentum.
For example, in the calculation of the cross section for $\gamma D\to pn\phi$~\cite{Laget:1994ba}, it was noted that if the $\phi$ interacts with the nucleus through two-gluon-exchange (Fig.~\ref{fig:Laget_hiddencolor} Left b) then, since coupling of a single gluon to a quark changes its color, it could transform a hidden-color configuration into a detectable state of 2 color-singlet nucleons, Fig.~\ref{fig:Laget_hiddencolor} Left a.
Estimates of the differential cross section for $D(\gamma,p\phi)n$ at high $-t$ for the quasi-free production from two-gluon-exchange are shown in Fig.~\ref{fig:Laget_hiddencolor} Right.  Calculations are done using the Paris potential~\cite{Lacombe:1981eg} and for a hidden color component of 0.1\% or 1.0\% using the wavefunction of Ref.~\cite{Yamauchi:1986cv}.  
Measurements at high momenta would distinguish between these scenarios.
A six-quark hidden-color component of 0.15\% that describes the HERMES $b_1$ result~\cite{Miller:2013hla} cannot be ruled out by any existing observations.

One may look at higher and higher momenta, but the rates drop exponentially with $-t$.
The ability to separate $S$- and $D$-wave components from deuteron scattering through separation of the spin states is a significant advantage here.
By emphasizing the $S$-wave component, one will significantly decrease the amount of high momentum component. 
This can be seen in Fig.~\ref{fig:D_wavefunction} for momenta between 300 and 800 MeV/c.
The ability to decrease the expected high momentum component makes one more sensitive to unexpected components there.

The arguments above apply even more clearly to the $J/\psi$ since the mass scale is clearly separated from the momentum transfer scale and the object is significantly more compact.
The number of $J/\psi$ produced from deuterons in the experiment will be modest, on the order of 150 events~\cite{Pybus:2024ifi}.
The $J/\psi$ photoproduction near threshold has recently proven to be a fruitful observable to measure and our experiment will provide, even with limited statistics, a proof-of-principle for such study of hidden color.

\subsection{Vector Polarization Observables \label{sec:add_physics:vectorpol}}

While this is a proposal for tensor-polarized deuteron and the experimental set-up will be optimized for that purpose, vector-polarized target data are necessarily obtained concurrently, see Fig.~\ref{fig:triangle}.
To ensure that we can separate $m=0$ from $m=\pm1$, we must also be able to separate $m=+1$ from $m=-1$, see Sec.~\ref{sec:target}.  
The vector polarization will allow us to measure two beam-target double spin asymmetries. The first, the beam–target double-polarization photoproduction asymmetry $\mathbb{E}$ (also called the helicity asymmetry)~\cite{Barker:1975bp} will be measured over the full beam energy range of the experiment.  The second, the linear-beam longitudinal-target double-polarization photoproduction asymmetry $\mathbb{G}$~\cite{Barker:1975bp} will be measured in the energy range of the coherent peak, $\approx9$~GeV.

The availability of these quantities opens up two interesting studies: (1) production from a quasi-free neutron by selecting events with unobserved proton; and (2) production from an initial configuration with high relative momentum between the nucleons.
A first look at the helicity asymmetry will be available from the REGGE Experiment~\cite{Dalton:2020wdv} 
but this proposal will have significantly higher integrated luminosity.
$\mathbb{E}$ has not yet been measured in the energy range of interest for this experiment.

\subsubsection{Quasi-free production from the neutron}

Selecting events without a proton detected in the Start Counter makes us sensitive to neutrons with small initial momentum in the nucleus.  This allows us to measure $\mathbb{E}$ and $\mathbb{G}$ on quasi-free neutrons for beam energies spanning from the high resonance region up to the Regge domain.
Benchmarking $\mathbb{E}$ and $\mathbb{G}$ will turn them into means to study high relative momentum components of the deuteron wavefunction, as described in the next section.
In addition, measurements from a quasi-free neutron can be used to study the isospin dependence of Regge phenomenology.
For example, $\gamma n \to \pi^- \Delta^{+}$ is an isospin rotation from $\gamma p \to \pi^- \Delta^{++}$, as is $\gamma n \to \pi^- p$ from $\gamma p \to \pi^+ n$.
Finally, double-polarization studies with a single meson recoiling from a $\Delta$ baryon allows us 
to study its spin-density matrix elements, which can then be used to constrain amplitudes for use in final states with multi-mesons recoiling from a $\Delta$~\cite{Afzal:2024dbo}.

\subsubsection{Correlated pair production and $D$-wave depolarization}

The $D$-wave component of the wavefunction has an effective polarization of the nucleons half that of the  $S$-wave and in the opposite direction.
This leads to ``$D$-wave depolarization", which is sensitive to the details of the deuteron wavefunction.
Thus, this opens the possibility of a direct and simple measurement of the degree of polarization of nucleons in the nucleus as a function of relative momentum and thereby mapping, using $\gamma D\to \rho pN$, the relative size of the $S$ and $D$-wave.
$\mathbb{E}$ and $\mathbb{G}$ will monitor the polarization of the struck nucleon as a function of the pair relative momentum.
The asymmetry here is simply used as an indicator of nucleon polarization with its unmodified value known from the E12-20-011 REGGE Experiment for proton and for neutron from quasi-free production in this experiment.

\section*{Acknowledgements}
We acknowledge helpful conversations with Alex Gnech, Misak Sargsian, Simon Sirca, Justin Stevens, Ted Rogers, Christian Weiss and Bo Yu.  
Bo Yu kindly provided data from his analysis of 2021 SRC-CT data.
Misak Sargsian kindly provided the model used for the coherent production of coherent vector mesons.

\newpage

\printbibliography{}

\end{document}